\documentclass[11pt,fullpage]{article} 

\usepackage{comment}
\usepackage{setspace}
\usepackage{graphicx}
\usepackage{amsfonts}
\usepackage{xcolor}
\usepackage{tikz}
\usepackage{slashbox}
\usepackage{varwidth}
\usepackage{caption}
\usepackage{subcaption}
\usepackage{authblk}
\usepackage{booktabs}
\usepackage{multirow}
\usepackage{enumerate}
\usepackage{url}

\usepackage[round,authoryear]{natbib}

\usepackage[english]{babel}
\usepackage[utf8]{inputenc}
\usetikzlibrary{arrows}
\usetikzlibrary{shapes}
\usepackage{amsmath}

\tikzset{
    *|/.style={
        to path={
            (perpendicular cs: horizontal line through={(\tikztostart)},
                                 vertical line through={(\tikztotarget)})
            -- (\tikztotarget) \tikztonodes
        }
    }
}

\begin{document}

\title{Pricing Barrier Options with DeepBSDEs\footnote{Self-archived author accepted manuscript version of the article published in Journal of Computational Finance, 
Volume 25, Number 4, Pages 1-25, DOI: \protect\url{https://doi.org/10.21314/JCF.2021.016}}.\footnote{Any opinions, findings, conclusions, and/or recommendations
expressed in this material are those of the authors and do not necessarily reflect the views of any past, current, or future employers (such as 
Wells Fargo Bank, N.A., its parent company, affiliates, and subsidiaries) of any of the authors.}} 
\author[1]{Narayan Ganesan}
\author[1]{Yajie Yu} 
\author[1]{Bernhard Hientzsch}
\affil[1]{Corporate Model Risk, Wells Fargo (as of March 2022)}
\date{}

\maketitle
\begin{abstract}
This paper presents a novel and direct approach to solve boundary and
final-value problems, corresponding to barrier options, using forward pathwise
deep learning and forward-backward stochastic differential equations (FBSDEs).
Barrier instruments are instruments that expire or transform into another
instrument if a barrier condition is satisfied before maturity; otherwise they
perform like the instrument without the barrier condition. In a PDE
formulation, this corresponds to adding boundary conditions to the final value
problem. The deep BSDE methods developed so far have not addressed
barrier/boundary conditions directly. We extend the pathwise forward deep BSDE
to the barrier condition case by adding nodes to the computational graph to explicitly
monitor the barrier conditions for each realization of the dynamics as well as
nodes that preserve the time, state variables, and trading strategy value at
barrier breach or at maturity otherwise. Given these additional nodes in the
computational graph, the forward loss function quantifies the replication of the
barrier or final payoff according to a chosen risk measure such as squared sum
of differences. The proposed method can handle any barrier condition in the
FBSDE set-up and any Dirichlet boundary conditions in the PDE set-up, both in
low and high dimensions.
\end{abstract}

\section{Introduction}
Deep Learning and Deep Neural Networks have been applied to numerically solve
high-dimensional nonlinear PDEs via the use of Forward-Backward Stochastic
Differential Equations or FBSDEs (see \citep{han2018solving}). In particular,
\citep{han2018solving} applied it to a quantitative finance pricing problem to
 price a combination of two Call options under differential rates (different
 lending and borrowing  interest rates), a nonlinear problem that was also studied in
\citep{mercurio2015bergman}. In related work, \citep{chan2019machine} and
\citep{raissi2018forward} also showed the applicability of deep learning in
solving  FBSDEs. The overview paper of DeepBSDE methods
\citep{hientzsch2019intro} introduces how FBSDE and certain deep learning methods
can be used to solve quantitative finance problems.

This paper presents a novel approach to solve boundary and final value problems
(corresponding to barrier options) with pathwise forward deep learning
approaches (``forward DeepBSDE''). The approach adds nodes to the computational
graph to explicitly monitor the barrier conditions for each realization of the
dynamics.
In case of barrier breach, the nodes record the time and underlying state
variables; otherwise they record the final time and value at maturity which is
used to determine the final payoff. The forward loss function then quantifies
the replication of the barrier or final payoff according to a chosen risk
measure such as squared sum of differences.

The simplest forms of barrier options are single-underlier or basket knock-in or
knock-out options in which the barrier condition only involves the current value
of the underlier or the basket and compares it to a given barrier level, which
is often a given constant. Let $S_t$ be that underlier or the value of the
basket. Examples shown in Table \ref{table:barriervariants} include the variants
of simple barrier Calls in this setting, for strike price ($K$), time to
maturity ($T$), a barrier positions $U_t$ and $L_t$ (possibly time dependent)
and the vanilla Call price $V(S_t,K,T-t)$. Put versions of these examples are
obtained by replacing Calls and Call prices by Puts and Put prices everywhere.

\begin{table}[h]
\small
\begin{tabular}{|r|p{1.08in}|p{1.08in}|p{1.08in}|p{1.08in}|}
  \hline
  Name & Barrier Condition & Knocked-in Instrument Rebate & Knocked-In
  Instrument Value & Final Payoff if not Breached \\
  \hline \hline 
  Up-and-Out Call & $S_t \geq U_t$ (Upper Barrier Position) & $G_t$  & $0$ &
  $\max(S_T-K,0)$  \\
  \hline 
  Down-and-Out Call & $S_t \leq L_t$ (Lower Barrier Position) & $G_t$  & $0$ &
  $\max(S_T-K,0)$  \\
  \hline 
  Up-and-In Call & $S_t \geq U_t$ (Upper Barrier Position) & $0$  & $V(S_t, K,
  T-t)$ & $0$  \\
  \hline 
  Down-and-In Call & $S_t \leq L_t$ (Lower Barrier Position) & $0$  & $V(S_t, K,
  T-t)$ & $0$  \\
  \hline
\end{tabular}
\caption{Summary of basic barrier Call instruments}
\label{table:barriervariants}
\end{table}

The simplest barrier options are those with barrier at some constant level which
is active for the entire life of the instrument. Two standard examples: A
standard Up-and-Out Call option with upper barrier $B$ will pay the
final Call payoff unless the underlier $S$ of the option was observed at a level
$S \geq B$ during the life of the option in which case it will pay a rebate $G$.
A standard Up-and-In Call option with upper barrier $B$ will pay the
final Call payoff only if the underlier $S$ of the option was observed at a
level $S \geq B$ during the life of the option and otherwise will pay a rebate
$G$.

Given a knock-in instrument, a model, and a valuation approach applied to that
model, the knocked-in instrument at barrier breach can be replaced by an
immediate payment of the value of the knocked-in instrument according to the
model without impacting the value of the knock-in instrument before barrier
breach.
In this way, one can restrict oneself to the case of knock-out instruments with rebates.

There are many instruments with barrier features in Quantitative Finance. There
are also many other applications in natural sciences, engineering, and economics
that involve bounded domains and boundaries in stochastic analysis and
stochastic processes. Similarly, PDE models in natural sciences and engineering
are often posed in bounded domains with boundary conditions imposed at the
boundary. The extensions presented in this paper to the forward DeepBSDE method
allow this new methodology to be applied to all these situations and
applications.

The rest of the paper is organized as follows: Section \ref{sec:deepbsdeintro}
presents a brief overview of Forward-Backward Stochastic Differential Equations
as applied to options pricing. Section \ref{sec:dnnsolution} discusses the
technique behind DeepBSDE approach to solving traditional option pricing problem
and a description of the approach to problems with barriers. Section
\ref{otherapproaches} discusses other approaches that are currently used or
proposed for pricing barrier options and outlines their limitations. Section
\ref{barriertrig} describes our approach, which employs barrier tracking
variables as a part of the DeepBSDE graph. Section \ref{sec:results} presents
the results of pricing barrier options in one and ten dimensions using this
approach. The paper concludes with some remarks.

\section{Problem Setup and Forward-Backward SDEs}
\label{sec:deepbsdeintro}

For a general introduction to FBSDE and their relations to nonlinear parabolic
PDE and/or quantitative finance, see \citep{el1997backward} and
\citep{perkowski2010}. \citep{han2018solving} presented the first application of
DeepBSDE method to the pricing of European options under differential rates.
\citep{hientzsch2019intro} provides an overview of the types of problems in
quantitative finance that can be solved using DeepBSDE and FBSDE. In this paper,
we will discuss only specific aspects of PDEs and FBSDEs relevant to our
setting.
 
We will first discuss how a semilinear parabolic PDE can give rise to a FBSDE.
A semilinear parabolic PDE is a PDE of the form:
\begin{eqnarray}
\frac{\partial u}{\partial t}(t,x) & + & Lu(t,x) + f(t,x,u(t,x), a^T(t,x)\nabla_x u(t,x))=0 \label{semilinparabolicpde} \\
Lu(t,x) & := &   \frac{1}{2} \mathbf{Tr}\left(a(t,x)a^T(t,x)\nabla^2_x u(t,x)\right) + \nabla_x u(t,x) b(t,x) \nonumber
\end{eqnarray}
defined on a domain $D\subset \mathbb{R}^d$ and $x\in D$. In quantitative
finance and other applications, these PDE  are often posed with terminal
conditions, $u(T,x) = g(T, x)$ for a given  function $g$.

Consider an It\^{o} process $X$ given by:
\begin{equation}
dX_t = b(t,X_t)dt + a(t,X_t)dW_t
\label{xprocess}
\end{equation}
where $X_t\in \mathbb{R}^d, dW_t\in \mathbb{R}^d, b(t,X)\in \mathbb{R}^d$ and
$a(t,X)\in \mathbb{R}^{d\times d}$,   with uncorrelated $dW_t$ and correlations
expressed through $a(t,X)$, defined on an appropriate filtration
$\mathcal{F}_t$.
For every $(t,x)$ we will consider a version of $X$ that starts (or arrives) at
time $t$ in $x$ and call it $X^{t,x}$.  
Let $Y_t$ be $u(t,X_t)$. Then $Y_t$  satisfies the Backward Stochastic
Differential  Equation
\begin{equation}
dY_t = -f(t,X_t, Y_t, a^T(t,X_t)\nabla u(t,X_t))dt + (\nabla u(t,X_t))^T a(t,X_t)
dW_t
\label{yprocess}
\end{equation}
along with the equation for $X_t$, satisfying (\ref{xprocess}). 
A terminal condition $u(T,x) = g(T, x)$ for the PDE translates into a terminal
condition for the BSDE $Y_T= g(T, X_T)$. 

With $\pi_t = \nabla_x u(t,X_t)$, the BSDE  (\ref{yprocess}) can be written 
\begin{equation}
-dY_t = f(t,X_t, Y_t, a^T(t,X_t)\pi_t) dt - \pi^T_t a(t,X_t) dW_t
\label{yzprocess}
\end{equation}

Equations (\ref{xprocess}) and (\ref{yzprocess}) together describe the dynamics
of solution $u(t,x)$ of PDE (\ref{semilinparabolicpde}) within the framework of
FBSDE. Similarly, one can prove there are functions $u$ and $\pi$ such that
$Y^{t,x}_s=u(s,X^{t,x}_s)$ and $\pi^{t,x}_s=\pi(s,X^{t,x}_s)$, that
$\pi(t,x)=\nabla_x u(t,x)$, and that $u$ then solves the  PDE
(\ref{semilinparabolicpde}) in the appropriate sense, see \citep{el1997backward}
and \citep{perkowski2010}.

So, instead of solving a PDE (\ref{semilinparabolicpde}) for $u$, we can solve
the FBSDE (\ref{yzprocess}) by finding stochastic processes $Y_t$ and $\pi_t$.
Instead of finding a process $\pi_t$ we can also try to find a function
$\pi(t,x)$ (or different functions for different $t$) such that
$\pi_t=\pi(t,X_t)$, as inspired by the expression for $\pi_t$ given above.

Notice that however the FBSDE formulation as written in  (\ref{xprocess}) and
(\ref{yzprocess}) is more general than the original PDE. For instance, the
terminal condition can be a random variable $G_T$ measurable as of
$\mathcal{F}_T$, depending on the paths of $X_t$ and $W_t$, rather than a
function of just the final value $X_T$.
Then, solution functions $u$ and portfolio functions $\pi$ would potentially
also have more general forms and more arguments and might also depend on the
state of additional processes that capture the path-dependency of the terminal
condition and allow the replication or risk management of the same. There is no
straightforward way to give equivalent PDE formulations at this level of
generality, in fact there might not exist a straightforward one. 

The above relation between PDE and FBSDE can be illustrated explicitly, for
instance in the case of risk-neutral valuation of simple derivatives which obey
the Black-Scholes-Merton equation
\begin{equation}
\frac{\partial u}{\partial t} +  r(t,x) x\frac{\partial u}{\partial
x}+\frac{\sigma^2(t,x)}{2}x^2\frac{\partial^2 u}{\partial x^2 } - r(t,x)
u(t,x)=0
\label{bsequation}
\end{equation}
with the underlying $X$, driven by an It\^{o} process, in the risk-neutral
measure \citep{book:shreve04}, under which $X$ follows 
\begin{equation}
dX = r(t,x) X dt + \sigma(t,x) X dW_t,
\label{Xprocess}
\end{equation}
where $r(t,x)$ is the risk-free rate and $\sigma(t,x)$ is the volatility of the
underlier.
Under this setup, for any function $y(t,x)$, by It\^{o} process rule, 
\begin{equation}
dy(t,x) = \left(\frac{\partial y}{\partial t} + r(t,x)x\frac{\partial
y}{\partial x}+\frac{\sigma(t,x)^2}{2}x^2\frac{\partial^2 y}{\partial x^2
}\right)dt + \sigma(t,x) x \frac{\partial y}{\partial x} dW_t
\label{itoruley}
\end{equation}
If $y(t,x)$ is to describe the behavior of the value of the derivative $u(t,x)$,
it has to obey equation (\ref{bsequation}), along with the terminal condition,
$y_T = g(T,X_T)$. 
Under this assumption (\ref{bsequation}), the first term on right side in
the parenthesis of equation(\ref{itoruley}) must obey,
\begin{equation}
\frac{\partial y}{\partial t} + r(t,x)x\frac{\partial y}{\partial
x}+\frac{\sigma(t,x)^2}{2}x^2\frac{\partial^2 y}{\partial x^2 } = r(t,x) y
\end{equation}
substituting for the term above in equation(\ref{itoruley}), leads to the
update equation for $y(t,x)$:
\begin{equation}
dy(t,x) = r(t,x) y(t,x)dt + \sigma(t,x) x \frac{\partial y}{\partial x} dW_t
\label{dy1}
\end{equation}
Now denoting $Y_t=y(t,X_t)$, $\pi(t)=\frac{\partial y}{\partial
x}(t,X_t)$, $a(t,X_t)=\sigma(t,X_t)X_t$, and $f(t,X_t,Y_t,Z_t)=-r(t,X_t)Y_t$,
this update equation corresponds to (\ref{yzprocess}). 

Here we are interested in the extension to PDE boundary conditions
corresponding to barrier options.
For PDEs, Dirichlet boundary conditions are imposed in addition to the PDE
(\ref{semilinparabolicpde}):
\begin{equation}
u(t,x) = g_B(t,x) \mbox{ for  }  (t,x) \in B
\end{equation}
corresponding to standard barrier conditions with barrier domain $B$.
Final value conditions can be included so that $g_B(T,x)=g(T,x)$ with the
previously defined $g$. We assume that this is done.

The function $g_B$ will encode the specific barrier option treated.
For a knock-out barrier Call option,  the barrier and final payoff is given by
\begin{equation}
g_B(t,X_t)=\left\{ 
\begin{array}{cc} 
\max(\mathrm{X_t}-K,0.0) & \mbox{ if } t=T \\
0.0 & \mbox{ if } t\leq T
\end{array}
\right. .
\end{equation}
A knock-in barrier Call option would be given by 
\begin{equation}
g_B(t,X_t)=\left\{ 
\begin{array}{cc} 
0.0 & \mbox{ if } t=T \\
V(X_t,K,T-t) & \mbox{ if } t\leq T
\end{array}
\right. 
\end{equation}
with $V(X_t,K,T-t)$ the vanilla Call price. 
For both examples, we assume that the barrier is not active at $T$.  
Note that $g_B$ is the same regardless of the level or number of
barriers. The first covers Up-and-Out, Down-and-Out, and
Double-Barrier-out Calls and similarly the second covers Up-and-In, Down-and-In,
and Double-Barrier-in Calls.

To include barrier conditions in the FBSDE formulation corresponding to the PDE,
the process $Y_t$ that we are trying to determine will follow the  FBSDE
(\ref{xprocess}) and  (\ref{yzprocess}) when $X_t$ is outside of $B$ while the
value of the process  $Y_t$ is directly given by $g_B(t,X_t)$ inside of $B$
(where maturity is  included in $B$).

More generally, define the barrier condition $C_t$ (eg: the barrier breach) as a
random variable with values $\mathtt{true}$ and $\mathtt{false}$ which is
measurable (computable) given the information about the dynamics $X_s,(s\leq
t)$. If a barrier domain $B$ in terms of state $(t,X_t)$ is given, then $C_t =
(t,X_t) \in B$. Similarly, define $\tau$ as the first time $(t,X_t)$ is within
$B$ respectively $C_t$ is true. The value of $Y_t$ at  $\tau$ can be given as a
random variable $G_\tau$ which is measurable (computable) given $X_s,(s\leq
\tau)$. If $g_B$ is given as above, $G_\tau=g_B(\tau,X_\tau)$.

Now a FBSDE with barrier condition can be formulated as a FBSDE with random
terminal time $\tau$. Given a stopping time $\tau$, for $t<\tau$, $Y_t$ follows
(\ref{yzprocess}):
\begin{equation}
-dY_t = f(t,X_t, Y_t, a^T(t,X_t)\pi_t) dt - \pi^T_t a(t,X_t) dW_t \nonumber
\end{equation} 
For times $t\geq \tau$, we assume the FBSDE is stopped
\begin{equation}
Y_t = G_\tau
\end{equation}
or in the barrier domain case,
\begin{equation}
Y_t = g_B(\tau,X_\tau).
\end{equation}
For completeness, we set $\pi_t=0$ for $t\geq \tau$. 

For general barrier conditions $C_t$, the stopping time is defined as
\begin{equation}
\tau = \inf \left\{ t \in [0,T]: C_t \mbox{ is true} \right\} \wedge T
\end{equation}
while for the barrier domain case, it is
\begin{equation}
\tau = \inf \left\{ t \in [0,T]: (t,X_t) \in B\right\} \wedge T
\end{equation}

Just as before, the stopped FBSDE formulation given here is much more general
than the PDE formulation presented earlier. For instance, the barrier trigger
and also the value of the option once barrier is triggered could be given in
terms of time averages, lookbacks to maximum or minimum, realized variance, etc.
Here, we will consider the barrier domain / boundary value setting.  

For this setting, \cite{lejay2002bsde} treats the case where either boundary
values are zero or they are given by the values of a bounded and H\"{o}lder
continuous function $\phi(t,x)$ that is defined everywhere and agrees with the
boundary values and final values in the appropriate domain. For the PDE, he
considers the weak solution defined through a variational form where the
boundary condition is treated as an essential condition (test and trial spaces
are $H^1_0$ on the complement of $B$ respectively the appropriate extension to
nonzero boundary values). He then proves that weak solutions of the PDE in that 
sense can be identified with solutions of the stopped FBSDE. In this setting and
under appropriate conditions, boundary conditions would be assumed pointwise.  
In a different context, \cite{kremsner2020deep} explores the relation between 
elliptical PDE with Dirichlet boundary conditions and stopped FBSDE
(corresponding to a setting where neither generator nor coefficients depend on
time). These Dirichlet boundary conditions are given by the values at the
boundary of a continuous bounded function $g$ defined on the domain of interest .
\cite{kremsner2020deep} shows that a classic solution of the PDE can be used to
construct a solution of the FBSDE and cites \cite{pardoux1998backward} which
shows that given a solution of the FBSDE, one obtains a continuous viscosity
solution of the elliptical PDE, and boundary conditions would be assumed
pointwise. In both cases settings, if the solution is smooth enough, the weak
respectively viscosity solution will also be a classical solution. Thus, under
these conditions, one can use FBSDE to find solutions of the PDE that will
assume the boundary conditions pointwise. We will not further consider the
theory for boundary value problems or stopped FBSDE for more general cases or
under weaker conditions in this paper.  

In the risk-neutral case as mentioned above, if the barrier option is given by a
barrier domain $B$ and a barrier function $g_B(t,x)$, one can characterize the
solution of the boundary value problem as
\begin{equation}
u(t,x) = E[g_B(\tau_T^{t,x}, X^{t,x}_{\tau_T^{t,x} }) e^{-\int_t^{\tau_T^{t,x}} r(s,X_s^{t,x})ds}]
\label{barrierexp1}
\end{equation}
where the expectation is taken with respect to the risk-neutral measure
\citep{book:shreve04} and the stopping time is here defined as
\begin{equation}
\tau_T^{t,x} = \inf \left\{ s \in [0,T]: (s,X^{t,x}_s) \in B\right\} \wedge T
\end{equation}
using the version $X^{t,x}_s$ of $X$ started in $x$ at time $t$.

For any general FBSDE problem as in Equations
(\ref{xprocess}) and (\ref{yzprocess}), applying a simple Euler-Maruyama
discretization for both $X_t$ and $Y_t$, we obtain
\begin{equation}
X_{t_{i+1}} = X_{t_i} + b(t_i,X_{t_i}) \Delta t_i + a(t_i,X_{t_i})
\Delta W^i \label{xupdate}
\end{equation}
\begin{equation}
Y_{t_{i+1}} = Y_{t_i} - f\left( t_i,X_{t_i},Y_{t_i},\pi_{t_i} \right) \Delta 
t_i + \pi^T_{t_i} a(t_i,X_{t_i}) \Delta W^i \label{yupdate}
\end{equation}
This can be used to time-step both $X_t$ and $Y_t$ forward. 

$X_t$ will be simulated forward in a pathwise sense. The equation for $Y$ can be
time-stepped forward in a pathwise sense or can be understood as an equation
that connects a value function for $Y$ at time $t_i$ with a value function of
$Y$ at time $t_{i+1}$, either in an exact or a least square sense (assuming that
the $\pi_{t_i}$ is given as a function or process). Under
appropriate conditions, one can solve the $Y$ equation for $Y_{t_i}$ in terms of
$Y_{t_{i+1}}$ and then time-step it also backward.
A value function can be fitted to pathwise values either as a post-processing
step or as an additional term in the appropriate loss function.
In the methods considered in this paper, we will time-step the $Y$ equation
pathwise forward (forward pathwise FBSDE methods).

\section{Solving BSDE with Deep Neural Networks}
\label{sec:dnnsolution}

We will first discuss application to European option pricing (see
\citep{han2018solving}). The FBSDE is first discretized in time as in
(\ref{xupdate}) and (\ref{yupdate}). The portfolio process $\pi_t$ is
represented by functions $p_n(X_{t_n})=\pi(t_n,X_{t_n})$, one for each time
$t_n$, and each function $p_n(X_{t_n})$ is given as a deep neural network. The
initial value of the portfolio at the fixed $X_0$ and the initial portfolio
composition $p_0$ are given as constants.

To generate a realization/training sample, one first generates a realization of
$X_{t_n}$ by (\ref{xupdate}) and then given the current parameter values for
$Y_0$ and all the $\pi_{t_n}=p_{n}$ networks, one computes $Y_{t_n}$ pathwise
and finally $Y_{T}=Y_{t_N}$ step by step by (\ref{yupdate}). The $L^2$ norm of the
replication error $Y_{t_N} - g(t_N,X_{t_N})$ is used as a loss function.

Instead of using a constant $X_0$, one can start with a randomly generated
$X_0$. Under those circumstances, one determines the parameters of the network
representing the function $Y_0(X_0)$ rather than a single value $Y_0$ (and also
a $\pi_0 = p_0$ network and its parameters). We have also implemented that
method. \citep{han2018solving} mention the forward approach for random $X_0$ as a
possibility on page 8509 but we are not aware of any implementation of this
method besides our own to the best of our knowledge. One can also use other risk
measures defined from the replication error in the optimization. Finally, the
loss function is optimized with stochastic optimization methods such as
mini-batch stochastic gradient descent combined with Adam optimizers or other
appropriate deep learning methods to determine the parameters of the $p_n$ and
$Y_0$ network. \\
Figure \ref{fig:dbsdenetwork} shows the computational graph for the forward
method for European options with random $X_0$.

\begin{figure}[h]
\centering
\resizebox{.95\linewidth}{!}{
\begin{tikzpicture}

\node[draw, fill=gray!20] (Y0) at (1.5,3) {$Y_0$ Network};
\node[draw, rounded corners, radius=1mm] (Y1) at (3.5,3) {$Y_1$};
\node[draw, rounded corners, radius=1mm] (Y2) at (6,3) {$Y_2$};
\node[text width=0.5cm] (Yi) at (7.5,3) {$\cdots$};
\node[draw, rounded corners, radius=1mm] (Yn_1) at (8.2,3) {$Y_{n-1}$};
\node[draw, rounded corners, radius=1mm] (Yn) at (11,3) {$Y_n$};

\node[draw, rounded corners, align=center] (Z0) at (2.5,2) {$\pi_0$ \\ Network};
\node[draw, rounded corners, align=center] (Z1) at (5,2) {$\pi_1$ \\ Network};
\node (Zi) at (7.5,2) {$\mathbf{\cdots}$};
\node[draw, rounded corners, align=center] (Zn_1) at (10,2) {$\pi_{n-1}$ \\ Network};

\node[draw] (X0) at (1.5,1) {$X_0$};
\node[draw] (X1) at (4,1) {$X_1$};
\node[text width=2cm] (Xi) at (6.5,1) {$\cdots$};
\node[draw] (XN_1) at (9,1) {$X_{n-1}$};
\node[draw] (XN) at (11.8,1) {$X_n$};

\node[draw] (W1) at (3.5,0) {$dW_1$};
\node[draw] (W2) at (6,0) {$dW_2$};
\node[text width=2cm] (Wi) at (8.2,0) {$\cdots$};
\node[draw] (Wn_1) at (8.2,0) {$dW_{n-1}$};
\node[draw] (Wn) at (11,0) {$dW_n$};

\node(po)[draw, ellipse, fill=red!8] at (14,2) {\begin{varwidth}{2cm}Payoff/Loss Function\end{varwidth}};

\draw[->](X0.north)-|(Z0.south);
\draw[->](X0.north)-|(Y0.south);
\draw[->](X0.east)to(X1.west);

\draw[->](X1.north)-|(Z1.south);
\draw[->](X1.east)to(Xi.west);

\draw[->](Xi.east)to(XN_1.west);
\draw[->](XN_1.north)-|(Zn_1.south);
\draw[->](XN_1.east)to(XN.west);

\draw[->](W1.north)to(Y1.south);
\draw[->](W2.north)to(Y2.south);
\draw[->](Wn_1.north)to(Yn_1.south);
\draw[->](Wn.north)to(Yn.south);

\draw[->](Y0.east)to(Y1.west);
\draw[->](Y1.east)to(Y2.west);
\draw[->](Y2.east)to(Yi.west);
\draw[->](Yn_1.east)to(Yn.west);

\draw[->](Y1.east)to(Y2.west);
\draw[->](Y2.east)to(Yi.west);
\draw[->](Yn_1.east)to(Yn.west);

\draw[->](Yn.east) to (po);
\draw[->](XN.east)to (po);

\path[every node/.style={font=\sffamily\small}]
(Z0.north) edge[bend left] node [right] {} (Y1.west)
(Z1.north) edge[bend left] node [right] {} (Y2.west)
(Zn_1.north) edge[bend left] node [right] {} (Yn.west);

\end{tikzpicture}}
\caption{Computational graph for forward DeepBSDE} \label{fig:dbsdenetwork}
\end{figure}


As seen earlier, in the time-continuous case, $\pi_t = \nabla_x u(t,X_t)$, which
in financial terms corresponds to delta-hedging according to some value function
$u$. In the case the FBSDE, time-continuous or not, is expressed as
minimization, the $\pi_t$ determined during the optimization does not
necessarily have to be equal to $\nabla_x u(t,X_t)$ although it often
approximates it to a certain extent.
In the time-discrete case, the analytical solution is no longer guaranteed to
minimize the replication error, but reflects a good benchmark and what one would
achieve by continuous delta-hedging with that value function. Later, we will
compare the hedging/replication strategy given by the optimized DNN with the one
implied by the analytical solution, for the barrier case.

We now present the fundamental idea for the barrier case and will leave the
implementation details to a later section.
The fundamental idea to extend the forward pathwise method to the barrier case
is that the barrier breach time and place (together with maturity) takes the place of
maturity as the first time and place at which the value of $Y_t$ is known and
any approximation of $Y$ should try to approximate that first known value well
regardless whether it was specified in the problem as a final value or a barrier
condition value.

The first time at which the dynamics $X_t$ enters the barrier domain $B$ or
satisfies the barrier condition $C_t$ is called barrier touch/breach time and
is denoted $\tau_T^{t,x}$ as introduced above (where maturity and final
values are included in the domain $B$ and the $g_B$).
 
Thus, the loss function to be minimized, according to some norm/risk masure 
(for instance, $L^2$ norm), is 
\begin{equation}
Y_{\tau_T^{t,x}} - g_B\left(\tau_T^{t,x},X_{\tau_T^{t,x}}\right)
\label{barrierloss}
\end{equation}

If one would try to learn functions for $Y$ at each time going forward rather
than work with pathwise values, this would couple all functions and values at
all time steps together across all paths and all of them would be trained from
the replication mismatch at maturity or barrier breach, leading to a more
complex, harder, and harder to train problem. Also, if one determines solution
functions, one needs a way to enforce the barrier values in the regions where
the dynamics enters the barriers, and an efficient operational global
characterization of the barrier domains. For backward pathwise or function based
methods, one would need alternative formulations, not the stopped
characterization, which only works forward in time until the barrier has been
first breached. We leave adaptations of such methods to the barrier case to
future publications and present here the forward pathwise method which does not
have such issues and is more straightforward.

\section{Other Approaches to Pricing Barrier Options}
\label{otherapproaches}
We briefly review some of the techniques used in practice to price such barrier options.

{\bf PDE Based Approaches:} Options pricing via the solution of PDE with finite
differences or finite elements or other standard approaches in high dimensions
(many underlying assets) poses difficulties due to extreme storage and
computational requirements to compute and store the values on a grid that covers
the domain. Accuracy and stability requirements often require larger grids than
what is achievable with given resources. In traditional PDE schemes, the PDE
grid is generated with sufficiently small spacing for finite difference methods
in region of interest that includes final time $T$ and barrier positions. The
finite difference time-stepping proceeds from the final time $T$ along the time
axis spanning a domain bounded on one or both sides by barriers and/or other
appropriate boundary conditions.
The time-stepping scheme is then applied to determine the values at grid points
going backward in time. PDE techniques are well studied and applied widely,
however their applicability is limited and is best suited to pricing derivatives
on fewer number of underliers due to issues with storage and computational
requirements caused by dimensionality. However, PDE techniques can model at
least some nonlinear pricing.

{\bf Monte-Carlo Based Approaches:} Barrier options in high dimensions are often
priced by Monte-Carlo approaches. This includes generating multiple independent
sample paths of the underliers in order to compute the realized payoff at
maturity or barrier breach along each path and discounting it to the present
time and averaging the payoffs to determine the value of the option. While
generating sample paths in high dimensions certainly increases in difficulty and
resource requirements with the dimension, these requirements do not depend
exponentially on the number of dimensions as the grid size in PDE. In the case
when there are rare events with high impact on the price (such as a knock-out
condition on a short or far out barrier for an otherwise high payoff),
simulation-based approaches such as Monte-Carlo need to sample those rare events
sufficiently well to approximate the price of the instrument well. One of the
main limitations of standard Monte-Carlo based approaches is that they can only
take into account risk-neutral discounting or other linear pricing approaches
and cannot be used to model nonlinear pricing such as the differential rates, in
which borrowing and lending attracts different interest rates
\citep{hientzsch2019intro}.

{\bf DeepBSDE Based Approaches:} DeepBSDE based approaches as originally
proposed in \citep{han2018solving} alleviate the problem of dimensionality by
converting the high-dimensional PDEs into Backward Stochastic Differential
Equations. The SDEs are then solved as a stochastic optimal control problem by
approximating the co-state and initial value of the control problem using Deep
Neural Networks.  Barrier options require the handling of the barrier boundary
conditions as they correspond to `knock-in' and `knock-out's at any time before
maturity and  such problems are not treated in \citep{han2018solving}. 

As previously mentioned, the following expression (\ref{barrierexp1}) holds in
the risk-neutral case:
\begin{equation}
u(t,x) = E[g_B(\tau_T^{t,x}, X^{t,x}_{\tau_T^{t,x} }) e^{-\int_t^{\tau_T^{t,x}}
r(s,X_s^{t,x})ds}] \nonumber
\end{equation}
where the expectation is taken with respect to the risk-neutral measure
\citep{book:shreve04}.  

Under this risk-neutral measure $X$ follows (\ref{Xprocess}) with functions as
defined in Section \ref{sec:deepbsdeintro}, $\tau_T^{t,x}$ denotes the earlier 
of barrier breach and maturity, and $g_B$ combines both final values and barrier
values, as defined in Section \ref{sec:dnnsolution}.
For knock-out options with zero rebate, $g_B(t,x)$ will be zero unless $t=T$. 

Therefore, (\ref{barrierexp1}) will in this case simplify to 
\begin{eqnarray}
u(t,x) & = & E[1_{\tau^{t,x}\geq T} g(T, X^{t,x}_{T}) e^{-\int_t^{T} r(s,X_s^{t,x})ds}] \nonumber \\
& = & E[P\left({\tau^{t,x}\geq T}\right) g(T, X^{t,x}_{T}) e^{-\int_t^{T} r(s,X_s^{t,x})ds}]
\label{barrierexp2}
\end{eqnarray}

This corresponds to an instrument with final payoff 
\begin{equation}
P\left({\tau^{t,x}\geq T}\right) g(T, X^{t,x}_{T}),
\end{equation}
which is measurable as of time $T$ knowing $X$ up to that time $T$.
In the case that $P\left({\tau^{t,x}\geq T}\right)$ can be written as a 
(relatively simple and explicit) function of $X_t=x$ and $X_T$, this will give
a (different) final value problem for each $t$ and $x$.
Informally, the final payoff function is adapted via
a Brownian bridge based method to include the probability of breaching the
barrier before option maturity.

In recent work, \citep{yu2019deep} have approached the Barrier option pricing
as described by the above formula.
The computation of probability of breach $P(\tau^{t,x} \leq T)$ 
as a function of $x$, $t$, $X_T$, and $T$
is analytically tractable for constant barriers and simple risk models 
(constant drift and volatility of the underlying assets), with the help of
results from {\em Brownian bridge probabilities}. Once $x$ and $t$ are fixed, 
these are standard final value problems that can be solved with 
the DeepBSDE methods for European options, as demonstrated in
\citep{yu2019deep}.

A slight generalization can be derived along similar lines for the case in which
touching the barrier does not lead to an immediate rebate but only to a changed
final payoff, which we denote by $g_{Br}(T,x)$. Proceeding similarly to above,
one obtains an expectation that can be interpreted as an European option on the
final payoff
\begin{equation}
P\left({\tau^{t,x}\geq T}\right) g(T, X^{t,x}_{T}) + P\left({\tau^{t,x}<T}\right) g_{Br}(T, X^{t,x}_{T})
\end{equation}

However, in general, time varying barriers and stochastic risk-models
(stochastic interest-rate and volatility of underliers) require Monte-Carlo
based approaches to compute the probability. For instance for a single barrier
with different levels in multiple time-periods, the probability computation can
be extended as follows: \\
Consider an Up-and-Out barrier option with piecewise constant barriers with $M$
values within $0\leq t \leq T$, defined by increasing times $t_0=0$,
$t_i$,\ldots, $t_{M}=T$:
\begin{equation} B(t) = \{B_i \mbox{ for } t_i\leq t < t_{i+1}\}
\end{equation} 

The total probability of no-breach can be computed as product of probabilities
of no-breach within each time period, informally written as:
 \begin{eqnarray} 
  P(\tau^{t,x} \geq T) & = &
 \mbox{P}_{\mbox{NoBreach}}(t_0,X(t_0);T,X(T)) \nonumber \\
  & = & \prod_{i=0}^{M-1}
 \mbox{P}_{\mbox{NoBreach}}(t_i,X(t_i);t_{i+1},X(t_{i+1}))  
\end{eqnarray}
where each factor can be computed via Brownian bridge for simple
 risk-models.
However, now one needs to integrate over and/or otherwise sample over the
intermediate positions at intermediate times to obtain the barrier probabilities
as a function of only initial and final risk factor value. 
Some other popular varieties of barrier options including multiple barriers (Up
and Down barriers), interacting barriers or barrier levels specified as a
function of past asset price (maximum drawdown) require more involved
Monte-Carlo approaches to compute the probability.
Monte-Carlo approaches in higher dimensions to compute these probabilities are
computationally expensive and incur higher costs than pricing the option in
higher dimensions as outlined above (and therefore will not lead to an
efficient method to compute barrier option values through final value problems).
 
In this work, we propose an original approach using DeepBSDEs that explicitly
monitors whether the underliers breach the barrier before maturity and record
the state of the barrier option at every time step via nodes in the
computational graph, which is then used to determine the appropriate payoff
conditions. The technique proposed here is applicable to solving general
semi-linear parabolic PDEs or FBSDE in any discipline (in addition to
quantitative finance) in which boundary conditions are specified.
To the best of our knowledge there is no prior work to handle boundary
conditions explicitly in the context of DeepBSDEs, which is important in solving
a variety of PDEs and FBSDE with standard and non-standard boundary conditions.

During the publication of this paper, we were made aware of
\cite{kremsner2020deep}, where a similar pathwise forward DeepBSDE approach is
used to solve a FBSDE with random terminal time that corresponds to an
elliptical boundary value problem and infinite horizon control problems. Unlike
them, we handle the parabolic case and our method allows barrier conditions and
options that do not correspond to a single barrier domain.

\section{Deep Neural Network with Barrier Triggers}
\label{barriertrig}

As mentioned before, we need to keep track of variables that detect whether 
the barrier was breached (the barrier condition satisfied) and preserve the 
value of $X$, $t$, and $Y$ at the time of the first barrier breach. In general,
we will call these ``conditional'' variables, or ``conditional'' tensors (since variables 
are represented as and called tensors in TensorFlow). 

In the time-continuous setting, we need the values of $\tau^{t,x}_T$, 
$X_{\tau^{t,x}_T}$, and $Y_{\tau^{t,x}_T}$ to compute the loss function
(\ref{barrierloss}) or some other risk measure that evaluates how the trading 
strategy replicates the appropriate payoff.  In the time-discrete case, we
need to keep track of time-discrete counterparts, which we call 
$\mathbf{tFP}$ ($t$ for payoff), $\mathbf{XFP}$ ($X$ for payoff), 
and $\mathbf{YFP}$ ($Y$ for payoff). To write updates for these variables, 
we need $\mathbf{XTrig}$ that turns from false (0.0) to true (1.0) once 
the barrier has been breached. For barrier condition functions $C_t(t,X_t)$, 
this can be defined as 
\begin{equation}
\mathbf{XTrig}_i = \left\{ 
\begin{array}{cc} 
\mathbf{XTrig}_{i-1} & \mbox{ if } \mathbf{XTrig}_{i-1} \\
C_{t_i}(t_i,X_{t_i}) & \mbox{ else}
\end{array}
\right. .
\end{equation}
For a problem with a single upper barrier at level $U$ active during the entire 
time to maturity, this would read
\begin{equation}
\mathbf{XTrig}_i = \left\{ 
\begin{array}{cc} 
\mathbf{XTrig}_{i-1} & \mbox{ if } \mathbf{XTrig}_{i-1} \\
X_{t_i} \geq U & \mbox{ else}
\end{array}
\right. .
\end{equation}
$\mathbf{XTrig}$ would be appropriately initialized depending on whether the 
barrier condition will be checked at time $t_0$ and could possibly be true there
or whether it will only be checked starting at the next time step. 

Having defined $\mathbf{XTrig}$, we can define $\mathbf{tFP}$, $\mathbf{XFP}$, 
and $\mathbf{YFP}$ as follows:
\begin{eqnarray}
\mathrm{tFP}_{i} &=& \mathrm{t_i}\times (1.0-\mathrm{XTrig}_{i-1}) +
\mathrm{tFP}_{i-1}\times\mathrm{XTrig}_{i-1} \\
\mathrm{XFP}_{i} &= & X_{t_i}\times (1.0-\mathrm{XTrig}_{i-1}) +
\mathrm{XFP}_{i-1}\times\mathrm{XTrig}_{i-1} \\
\mathrm{YFP}_{i} &= & Y_{t_i} \times (1.0 - \mathrm{XTrig}_{i-1}) +
\mathrm{YFP}_{i-1}\times \mathrm{XTrig}_{i-1} 
\end{eqnarray}
and initialized with $tFP_{0}=t_{0}$,  $XFP_{0}=X_{t_{0}}$, and $YFP_{0}=Y_{t_0}$.
These update equations keep copying the underlying $X$, $Y$, and $t$ until the barrier
is hit and then stop at the next step, so that the values of the corresponding
variables at barrier breach stay in the *FP versions. If the barrier is never 
breached, the values of $t$, $X$, and $Y$ at $t=T=t_N$ will be in $\mathbf{tFP}_N=t_N$, 
$\mathbf{XFP}_N=X_{t_N}$, and $\mathbf{YFP}_N=Y_{t_N}$.
Thus, $\mathbf{tFP}_N$, $\mathbf{XFP}_N$, 
and $\mathbf{YFP}_N$ are the appropriate time-discrete analogues of 
$\tau^{t,x}_T$,  $X_{\tau^{t,x}_T}$, and $Y_{\tau^{t,x}_T}$, and the 
loss function for the computational graph implementing this time-discrete 
approach would be 
\begin{equation}
\left\| \mathbf{YFP}_N - g_B\left(\mathbf{tFP}_N,\mathbf{XFP}_N\right) \right\|^2
\label{tdbarrierloss}
\end{equation}
or any other appropriate risk measure. 

Notice also that time-discretization leads to well-known biases in barrier options. 
If approximation of a certain time-continuous problem with continuously enforced
barriers is desired, there are approaches such as barrier level correction (solving 
the time-discrete problem with appropriately shifted barrier levels) that will minimize
such bias and lead to faster convergence to the solution of the time-continuous 
problem.
 
Figure \ref{fig:barriernetwork} shows the computational graph for the approach
just described. The nodes that have been added or changed for the
barrier case are shown as color shaded nodes and the nodes from the original
forward DeepBSDE computational graph are shown unshaded and enclosed in a dashed box. As discussed above,
these additional tensors are needed to preserve the state at barrier breach or
maturity and serve to constrain the $Y_{t_i}$ appropriately so that it follows
the FBSDE if not in the barrier domain but approximates the given barrier domain
value once it touches the barrier domain.

\definecolor{columbiablue}{rgb}{0.61, 0.87, 1.0}
\definecolor{classicrose}{rgb}{0.98, 0.8, 0.91}
\definecolor{lightgray}{rgb}{0.83, 0.83, 0.83}

\begin{figure}
\centering
\resizebox{.95\linewidth}{!}{
\begin{tikzpicture}

\node[draw] (Z0) at (2.5,0) {$\pi_0$ Network};
\node[draw] (Z1) at (5,0) {$\pi_1$ Network};
\node (Zi) at (7.5,0) {$\mathbf{\cdots}$};
\node[draw] (Zn_1) at (10,0) {$\pi_{n-1}$ Network};

\node[draw] (Y1) at (3.2,1) {$Y_1$};
\node[draw] (Y2) at (5.7,1) {$Y_2$};
\node[text width=2cm] (Yi) at (8.2,1) {$\cdots$};
\node[draw] (Yn) at (10.7,1) {$Y_n$};

\node[draw, fill=lightgray] (tfpinit) at (0,8) {$\mbox{tFPinit}$};
\node[draw, fill=lightgray] (XTriginit) at (0,7) {$\mbox{XTriginit}$};
\node[draw, fill=lightgray, text width=24] (YFPinit) at (0,5.5) {\begin{align*}\mbox{YFP}_0\\ \mbox{XFP}_0\end{align*}};
\node[draw] (init) at (0,2) {init Network};
\node[draw] (Y0) at (0,1) {$Y_0$};

\node[draw, text width=20] (X0) at (1.9,2.5) {$X_0$};
\node(X0trig)[draw, ellipse,fill=blue!5, align=left]  at (2.1,7) {$\mbox{XTrig}_0$};
\node(t0fp)[draw,fill=red!5, align=left]  at (1.7,8) {$\mbox{tFP}_0$};
\node(YFP1)[draw,fill=green!5, text width=24]  at (3.2,5.5) {\begin{align*}\mbox{YFP}_1\\\mbox{XFP}_1\end{align*}};

\node[draw, text width=20] (X1) at (4.4,2.5) {$X_1$};
\node(X1trig)[draw, ellipse,fill=blue!5, align=left]  at (4.6,7) {$\mbox{XTrig}_1$};
\node(t1fp)[draw,fill=red!5]  at (4.2,8) {$\mbox{tFP}_1$};
\node(YFP2)[draw,fill=green!5, text width=24]  at (5.7,5.5) {\begin{align*}\mbox{YFP}_2\\\mbox{XFP}_2\end{align*}};

\node[text width=20] (Xi) at (7.4,2.5) {$\cdots$};
\node[text width=20] (XFPi) at (8.1,4) {$\cdots$};
\node(Xitrig)[text width=20]  at (7.7,7) {$\cdots$};
\node(tifp)[text width=45]  at (7.8,8) {$\cdots$};
\node(YFPi)[text width=20]  at (8.8,5.5) {$\cdots$};

\node[draw, text width=22] (Xn_1) at (9.4,2.5) {$X_{n-1}$};
\node(Xn_1trig)[draw, ellipse,fill=blue!5, align=left]  at (9.6,7) {$\mbox{XTrig}_{n-1}$};
\node(tn_1fp)[draw,fill=red!5]  at (9.2,8) {$\mbox{tFP}_{n-1}$};
\node(YFPn)[draw,fill=green!5, text width=25]  at (10.7,5.5) {\begin{align*}\mbox{YFP}_n\\\mbox{XFP}_n\end{align*}};

\node[draw, text width=20] (Xn) at (11.9,2.5) {$X_n$};
\node(tnfp)[draw,fill=red!5]  at (11.7,8) {$\mbox{tFP}_n$};

\node(po)[draw, ellipse, fill=red!8] at (14,5) {\begin{varwidth}{2cm}Payoff/Loss Function\end{varwidth}};

\draw[->] (Y0) to (Y1);
\draw[->] (Y1) to (Y2);
\draw[->] (Y2) to (Yi);

\draw[->] (Z0.north)-|(Y1);
\draw[->] (Z1.north)-|(Y2);
\draw[->] (Zn_1.north)-|(Yn);

\draw[->](tfpinit) to (t0fp);
\draw[->](XTriginit) to (X0trig);
\draw[->](YFPinit) to (YFP1);
\draw[->](init) to (Y0);

\draw[->](X0) to (X1);
\draw[->](X0) to (Y1);
\draw[->](X0.south) to ([xshift=-17]Z0.north);
\draw[->](X0) to[bend right] (init);

\draw[->](X0.north)-|(X0trig.south);
\draw[->](X0trig)to (t1fp);
\draw[->](X0trig)to (YFP1);
\draw[->](X0trig.east) to (X1trig.west);
\draw[->](t0fp) to (t1fp);
\draw[->](YFP1)to (YFP2);
\draw[->]([xshift=3]Y1.north) to([xshift=3]YFP1.south);

\draw[->](X1) to (Xi);
\draw[->](X1.south)to ([xshift=-17]Z1.north);
\draw[->](X1) to (Y2);
\draw[->](X1) to (Y2);
\draw[->](X1) to (YFP1);
\draw[->](X1.north)-|(X1trig);
\draw[->](X1trig)to (tifp);
\draw[->](X1trig)to (YFP2);
\draw[->](X1trig.east) to (Xitrig.west);
\draw[->](t1fp) to (tifp);
\draw[->](YFP2)to (YFPi);
\draw[->]([xshift=3]Y2.north) to([xshift=3]YFP2.south);

\draw[->](Xn_1.north)-|(Xn_1trig);
\draw[->](Xn_1.south)to ([xshift=-17]Zn_1.north);
\draw[->](Xn_1) to (Yn);
\draw[->](Xn_1) to (Xn);
\draw[->](Xn_1trig)to (YFPn);
\draw[->](Xn_1trig)to (tnfp);
\draw[->](tn_1fp)to (tnfp);
\draw[->]([xshift=3]Yn.north) to([xshift=3]YFPn.south);
\draw[->](YFPn.east) to (po);
\draw[->](tnfp.east)to (po);
\draw[->](Xn) to (YFPn);

\draw[gray,dashed] (-1.5,-0.5) rectangle (15,3) node[pos=.9] {\parbox[b][6cm]{2cm}{Original DeepBSDE Network}};

\end{tikzpicture}}
\caption{Computational graph for forward DeepBSDE with barrier
triggers}\label{fig:barriernetwork}
\end{figure}


\subsection{Behavior of the Conditional Barrier Tensors}

\begin{figure}[htpb!]
    \centering \includegraphics[width=.95\linewidth]{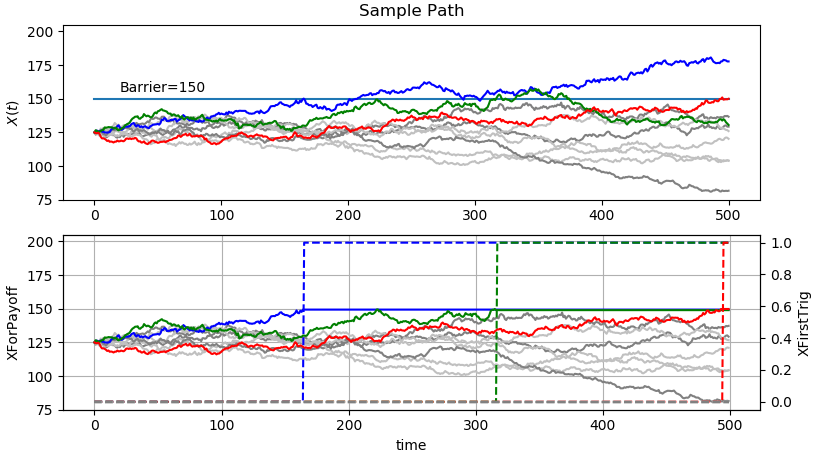}  
    \caption{Behavior of conditional barrier triggers}
    \label{xtrig}
\end{figure}

Figure \ref{xtrig} shows the values of these barrier breach tracking variables
for a case with 500 time steps, $X_0=125$ and an upper barrier at level 150, for
the same parameters of the $X$ dynamics as defined in Table \ref{table:rfmodel},
in the next section. The upper panel shows several sample paths of $X_{t_i}$ while the lower panel
shows the corresponding realizations of $\mathbf{XTrig}_i$ (scale on right axis) and $\mathbf{XFP}_i$ (scale on the left axis) for the same sample paths. 

The $\mathbf{XFP}$ stays at the value (``has been stopped'') it took at the first time when it breached the barrier level as shown for the sample paths highlighted (in color online). If the barrier is never touched, $\mathbf{XFP}$ is just $X$. The corresponding dashed lines, $\mathbf{XTrig}$, take the value 1.0 (true) at the time of barrier breach for the highlighted sample paths and otherwise stays at 0.0.

In the case with several barriers or more complicated barrier domains and conditions,
$\mathbf{tFP}_N$ and $\mathbf{XFP}_N$ would identify the time and place the
barrier was hit and thereby identify the barrier.

\section{Results}
\label{sec:results}

We perform testing in one and ten dimensions for Up-and-Out Calls on a
single underlier or a geometric basket of underliers following Black-Scholes
models. 

For an one-dimensional Black-Scholes model with constant coefficients $r$ and
$\sigma$, the price of an Up-and-Out Call with maturity $T$ at time $t$ and
$S_t$ with time-to-maturity $\tau=T-t$ can be computed as
follows\footnote{This is the result given, for instance,  in  
chapter 11 of \citep{nprivaultonline2021}, written in a more compact form.}:

\begin{eqnarray}
\delta^\tau_\pm (s) & = & \frac{1}{\sigma \sqrt{\tau}} \left( \log s + \left(r
\pm \frac{\sigma^2}{2}\right) \tau \right) \\
F_\pm (s) & = & \Phi\left(\delta^\tau_\pm\left(\frac{B}{K}s\right)\right) -
\Phi\left(\delta^\tau_\pm\left(s\right)\right) \\
G_\pm (s) & = & F_\pm(s) - \left(\frac{1}{s}\right)^{\pm 1 +
\frac{2r}{\sigma^2}} F_\pm\left(\frac{1}{s}\right) \\
\mbox{OUC}(t,S_t) & = & 1_{M^t_0<B} \left(S_t G_+\left(\frac{S_t}{B}\right) -e
^{-\tau r} K G_-\left(\frac{S_t}{B}\right)\right)
\end{eqnarray}
with $M^t_0=\sup_{s \in (0,t]} S_s$ being the minimum so far and $\Phi$ being
the cumulative density function of a standard normal. 

Assuming that there are $n$ correlated Black-Scholes
models $\frac{dS_i}{S_i}=\mu_i dt + \sigma_i dW_i$  with $dW_i dW_j = \rho_{ij} dt$,
one can easily show that the geometric basket  $GB(t) = \Pi^n_{i=1}
S_i(t)^{\alpha_i}$ will follow $\frac{dGB}{GB} = \bar{\mu}dt + \bar{\sigma}dW$ with 

\begin{equation}
\bar{\sigma} = \sqrt{\sum^n_{i,j=1} \alpha_i \alpha_j \rho_{ij} \sigma_i
\sigma_j} \qquad \bar{\mu} = \frac{\bar{\sigma}^2}{2} + \sum_{i=1}^n \alpha_i
\left(\mu_i -\frac{\sigma^2_i}{2} \right)
\end{equation}

We thus can analytically price an Up-and-Out Call on a geometric basket in an
arbitrary number of dimensions. 

If barrier breach is only monitored at discrete, regular times (such as
during MC simulation at regular intervals), \citep{kou2003pricing} shows that
the pricing impact of discrete observation can be approximated by shifting
the barrier away from spot by a certain multiplicative barrier correction, 
$e^{\beta\sigma\sqrt{\Delta T}}$ for $\Delta T$ being the size of the time step,
$\beta= -\zeta(\frac{1}{2})/\sqrt{2\pi}$ with $\zeta$ the Riemann zeta function,
and then pricing a barrier option with continuous observation with that barrier
level.

We use the parameters of the dynamics and instrument parameters as given in 
Table \ref{table:rfmodel}, along with initial value for the underliers chosen
uniformly between 49 and 151. For the ten-dimensional case, we use ten 
uncorrelated such models for the components and we use the geometric mean 
of the underliers raised to 1.05 as a basket underlier. The final payoff and the
barrier is defined in terms of that basket underlier. We note that Up-and-Out
Calls without rebate have a discontinuous $g_B$ function and are known
to be challenging, especially close to the corner of discontinuity.  

As for the FBSDE generator, we use the generator $f(t,X_t,Y_t,Z_t)=-r(t,X_t)Y_t$
for the risk-neutral (discounting-only) case, as shown in section 2. However,
any generator such as a generator for differential rates or other nonlinear
pricing settings could be easily used instead.

We use softplus as a (differentiable) activation function. We train separate
networks for the strategy function $\pi_t$ at different times $t$ and a separate
network for the $Y_0$ function. We tested several combinations of number of
hidden layers and of neurons in each layer.
More neurons on fewer layers performed better and so we used 20 neurons in 2
hidden layers for the one-dimensional case and 40 neurons in 2 hidden layers for
the ten-dimensional case.
Using more neurons than that in each layer did not seem to improve results
materially and neither did using more layers. We also tested a ResNet
architecture as in \citep{cyr2019robust} and observed similar but not better
results.\footnote{Note that \citep{cyr2019robust} considers approximation
problems while here we solve stochastic control problems optimizing over an
entire computational graph.}

We used Adam optimizer based on mini-batch gradient descent with various
mini-batch sizes and report results with mini-batches of size 1024. Larger
mini-batch sizes lead to better approximation of the loss function. However, a
mini-batch size of 1024 already leads to good visual approximation of the
analytical solution in appropriate settings.
We report 20,000 mini-batch steps but results for 10,000 mini-batches look
similar. We use exponentially decaying learning rate, starting at 0.01 and
decaying by 0.95 every 1,000 mini-batches.
Running the forward pathwise DeepBSDE method for barriers takes similar time as
the corresponding forward pathwise DeepBSDE method for instruments without
barriers. We have not tried to agressively optimize the learning rate schedule
or other features of the methods. Even so, the results for ten-dimensional
example are encouraging and we obtained good representations of solution and
strategy.
We will leave further performance improvements through different architectures
or algorithmic choices to future work.

\begin{table}
\centering
\begin{tabular}{|r|l|l|l|l|}
  \hline
    $\sigma$ & $r$ & $T$ & $B$ (Barrier Position) & $K$ (Strike) \\
  \hline \hline 
  0.2 &0.05 & 0.5 Years &150.0 & 100.0 \\ 
  \hline
\end{tabular}
\caption{Parameters for the $X$ dynamics, the generator, and the instruments}
\label{table:rfmodel}
\end{table}

\begin{figure}
\begin{subfigure}{.5\textwidth}
  \centering
  \includegraphics[width=.9\linewidth,height=.7\linewidth]{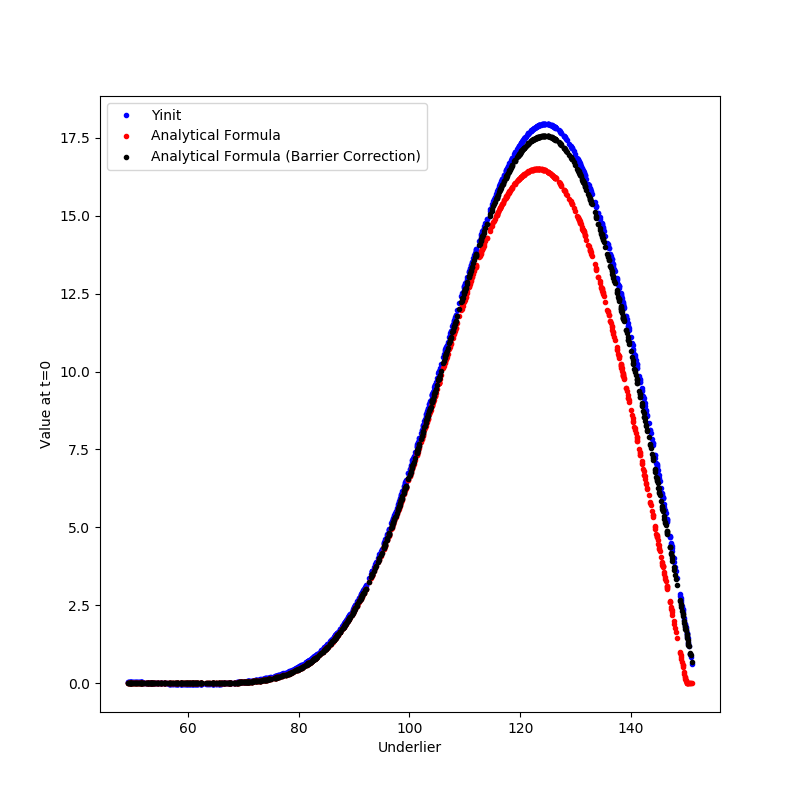}  
  \caption{50 time steps.}
  \label{onedim50ts}
\end{subfigure}
\begin{subfigure}{.5\textwidth}
  \centering
  \includegraphics[width=.9\linewidth,height=0.7\linewidth]{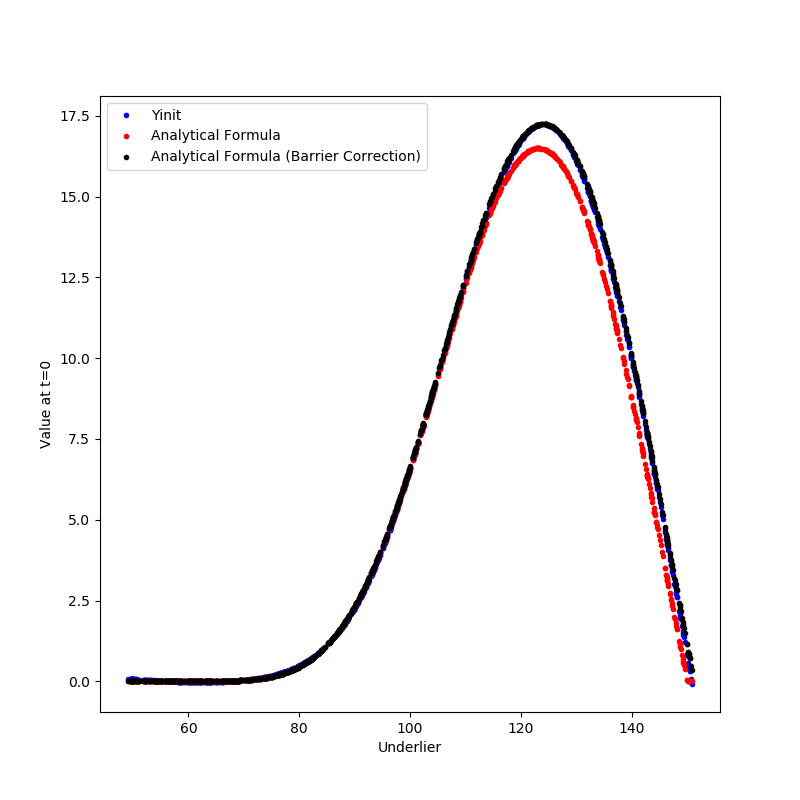}  
  \caption{100 time steps.}
  \label{onedim100ts}
\end{subfigure}
\caption[DeepBSDE for 1D Up-and-Out Call - value]{DeepBSDE value for
one-dimensional Up-and-Out Call with different number of time steps (20 neurons in 2 hidden layers).}
\end{figure}

In Figures \ref{onedim50ts} and \ref{onedim100ts} we show the DeepBSDE solution
$Y_0(X_0)$ (denoted Yinit here and in following figures) for 50 and 100 time
steps against the analytical solution with the original barrier position and the
continuity corrected barrier position. While there are still visual differences
to the continuity corrected analytical solution for 50 time steps, the solution
for 100 time steps agrees visually with the analytical solution where the
barrier position is corrected for the discrete monitoring of the barrier in the
simulation. We see that with increasing number of time steps, the continuity
corrected and the original analytical solution get closer and the DeepBSDE
solution approximates the continuity corrected analytical solution.

\begin{figure}
\begin{subfigure}{.5\textwidth}
  \centering
  \includegraphics[width=.9\linewidth,height=.7\linewidth]{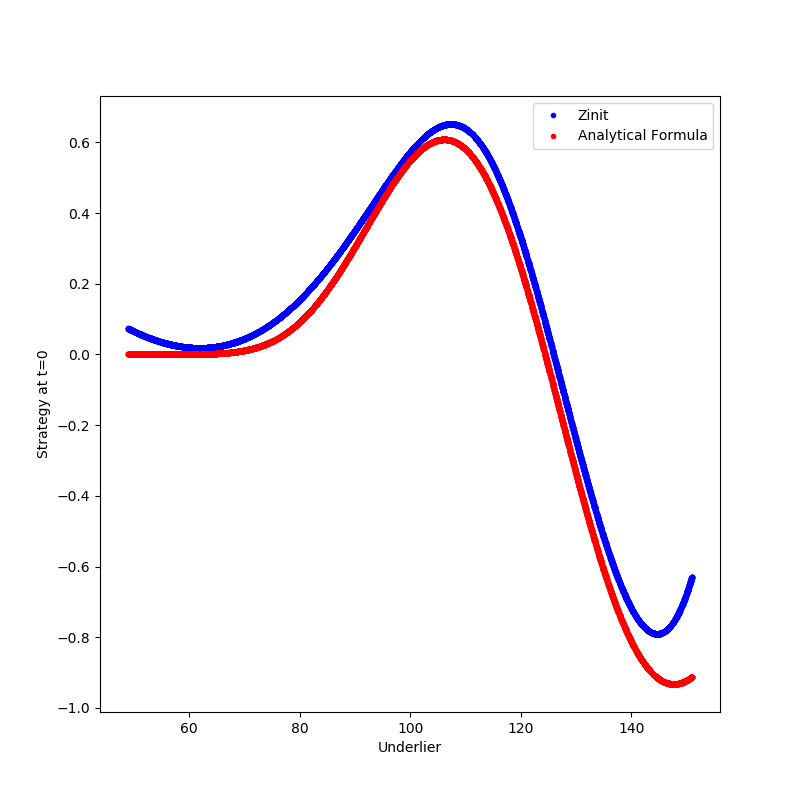}  
  \caption{50 time steps.}
  \label{onedim50tsdelta}
\end{subfigure}
\begin{subfigure}{.5\textwidth}
  \centering
  \includegraphics[width=.9\linewidth,height=0.7\linewidth]{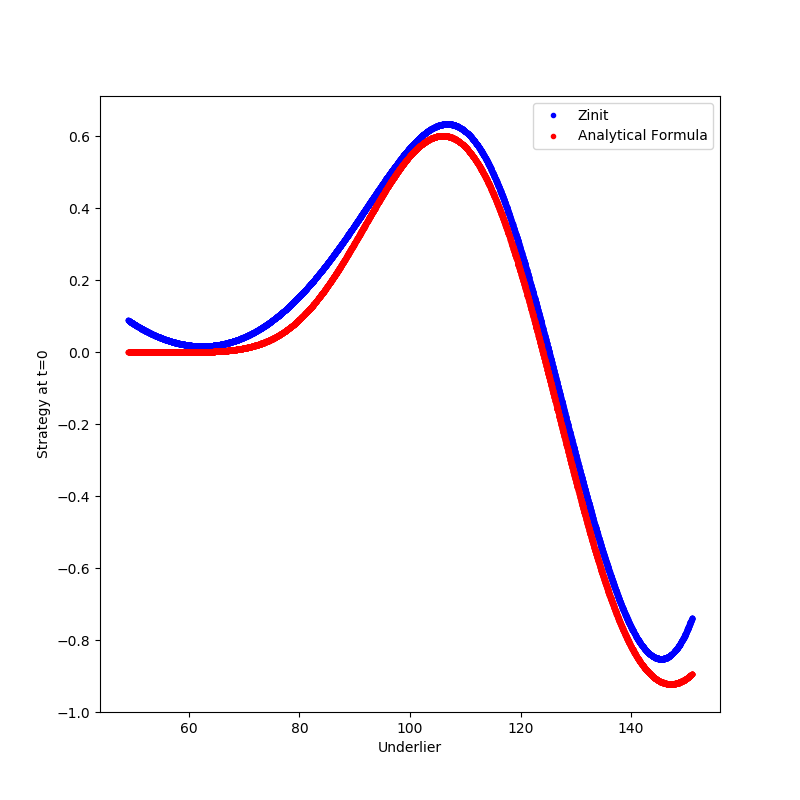}  
  \caption{100 time steps.}
  \label{onedim100tsdelta}
\end{subfigure}
\caption[DeepBSDE for 1D Up-and-Out Call - strategy]{DeepBSDE
strategy for one-dimensional Up-and-Out Call with different number of
time steps, on ten mini-batches (20 neurons in 2 hidden layers).}
\end{figure}

\begin{figure}
\begin{subfigure}{.5\textwidth}
  \centering
  \includegraphics[width=.9\linewidth,height=.7\linewidth]{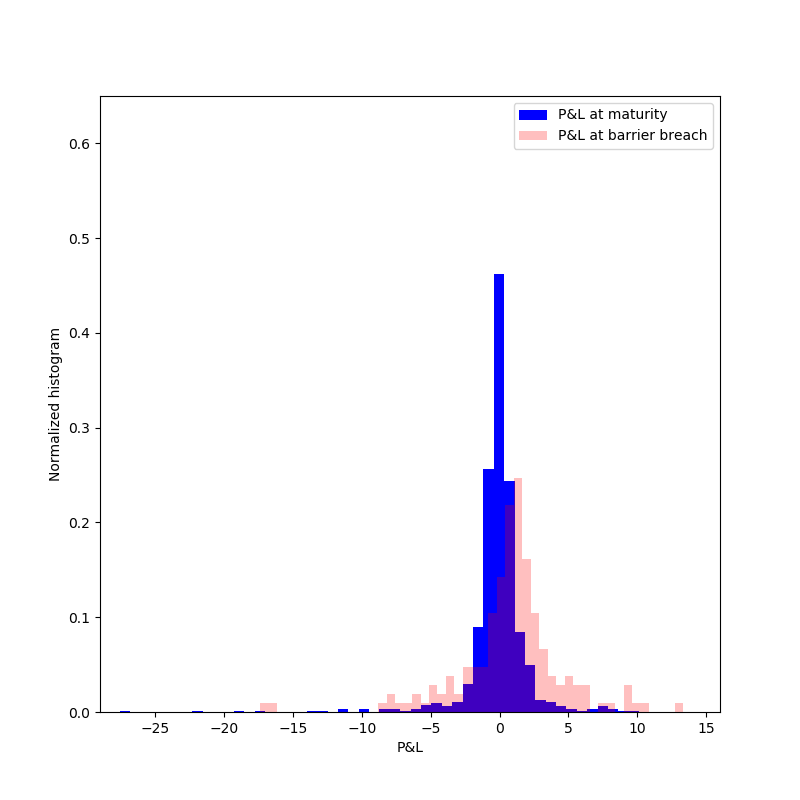}  
  \caption{DeepBSDE strategy.}
  \label{onedim50tsdeepbsdestrategy}
\end{subfigure}
\begin{subfigure}{.5\textwidth}
  \centering
  \includegraphics[width=.9\linewidth,height=0.7\linewidth]{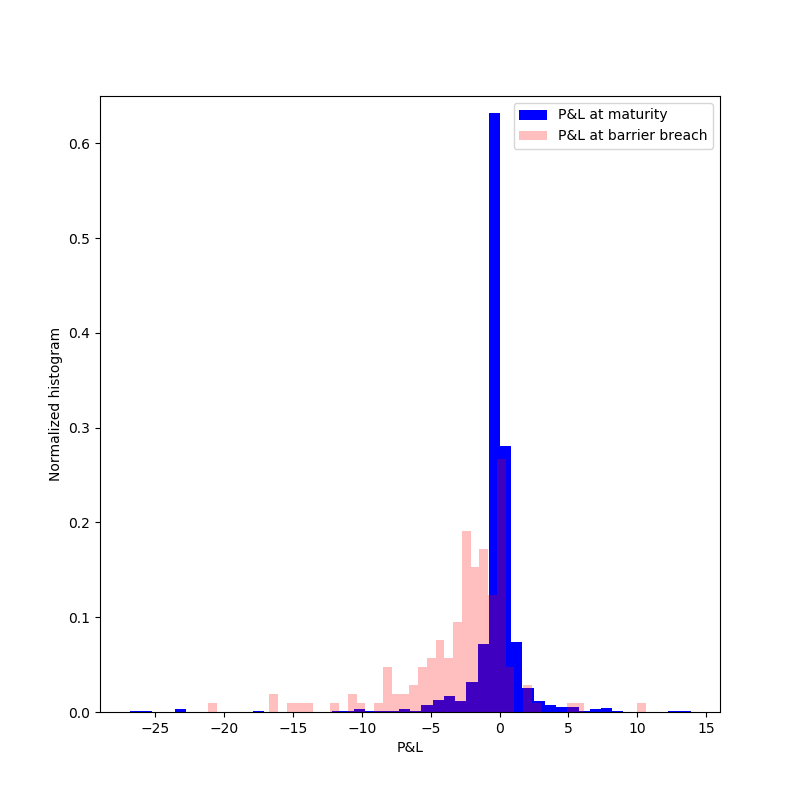}  
  \caption{Analytical delta.}
  \label{onedim50tsanalyticaldelta}
\end{subfigure}
\caption[P\&L for 1D Up-and-Out Call for DeepBSDE  vs. analytical
delta]{P\&L from DeepBSDE method vs. analytical delta from continuous-time problem. One-dimensional case,
50 time steps.}
\end{figure}

In Figures \ref{onedim50tsdelta} and \ref{onedim100tsdelta}, we compare the
hedging strategy at time zero as determined by DeepBSDE (called $\pi_0$ earlier
in the paper and denoted Zinit in the Figure) and the analytical delta for the
analytical solution with continuity corrected barrier position. We see that they
generally agree in shape and are reasonably close. For small values of the
underlier, the DeepBSDE strategy is not flat, but the strategy in that region
does not have a strong impact on the loss function or on the value
function.\footnote{If correct asymptotics are required, one can embed such
asymptotics in the network architecture for the strategy. We will leave such to
future work.} For larger values of the underlier closer to the barrier, the
DeepBSDE strategy is smaller in magnitude than the analytical strategy.
Recall however that the analytical strategy assumes continuous hedging rather
than discrete hedging and thus does not take into account that stock price might
have entered and remained in the barrier during the time step before rehedging
is possible. Looking at the later Figure \ref{onedim50tsanalyticaldelta}, we see
that the P\&L at barrier breach is skewed to negative P\&L. All other things
being equal, a less negative $\pi_t$ will increase the P\&L at barrier breach,
make it more symmetric around 0, and reduce the value of the loss function,
compared to using the more negative analytical delta close to the barrier.

\begin{figure}
\begin{subfigure}{.5\textwidth}
  \centering
  \includegraphics[width=.9\linewidth,height=.7\linewidth]{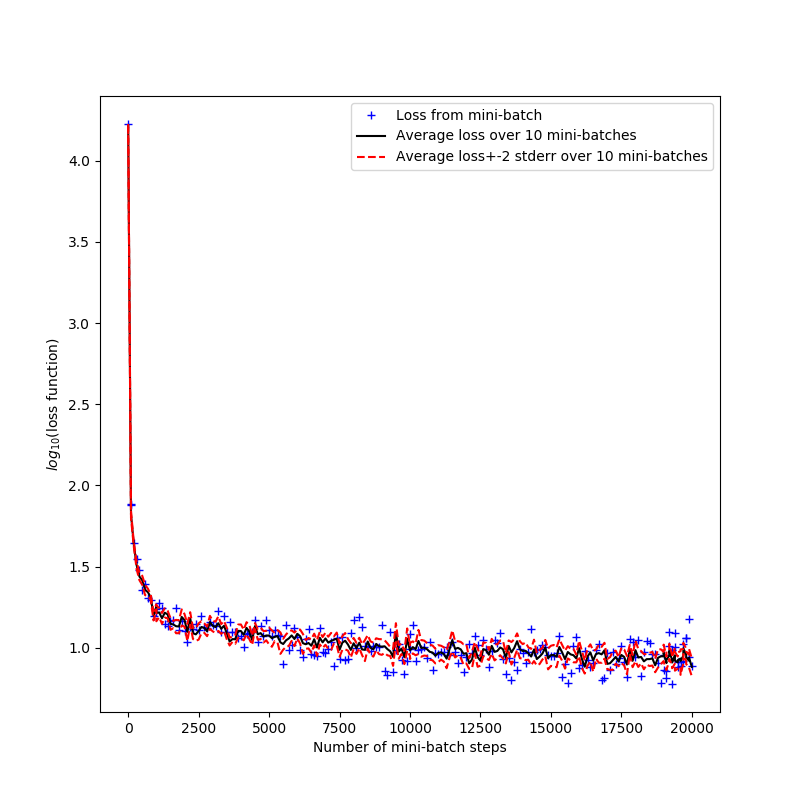}  
  \caption{50 time steps.}
  \label{onedim50tsloss}
\end{subfigure}
\begin{subfigure}{.5\textwidth}
  \centering
  \includegraphics[width=.9\linewidth,height=0.7\linewidth]{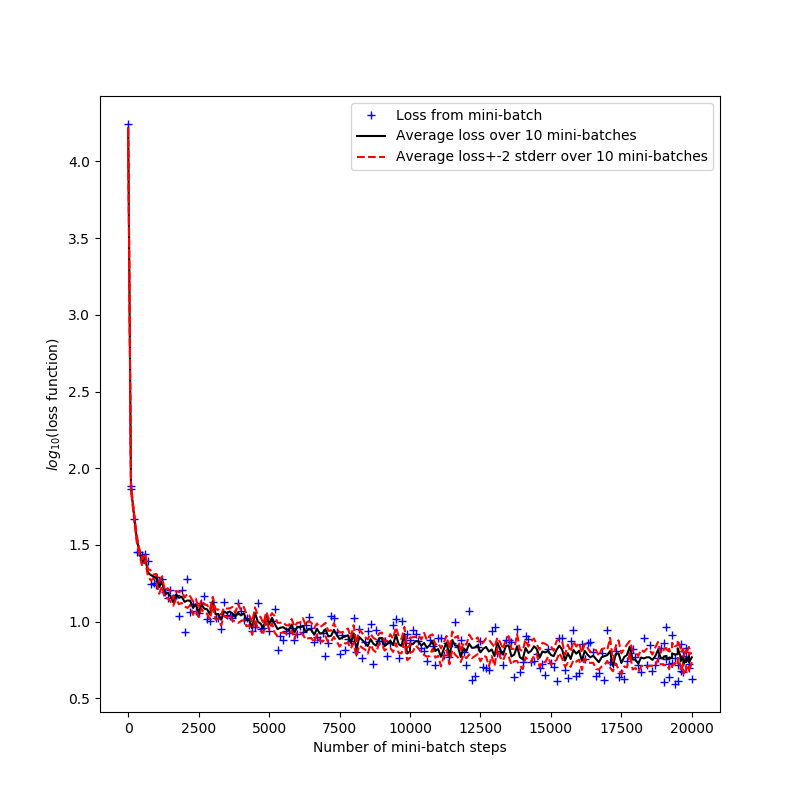}   
  \caption{100 time steps.}
  \label{onedim100tsloss}
\end{subfigure}
\caption[DeepBSDE for 1D Up-and-Out Call - loss]{DeepBSDE
loss function over number of mini-batch steps for one-dimensional Up-and-Out
Call with different number of time steps.}
\end{figure}

Figures \ref{onedim50tsloss} and \ref{onedim100tsloss} show the evolution of
the loss function during training for the cases with hedging at 50 and 100 time
steps. Since we are training one network for each time step, more time steps
mean more parameters to train so minima might be harder to find. More time
steps also mean more frequent hedging and therefore a chance to reduce P\&L
more. In the one-dimensional case, one can see that more time steps allow
DeepBSDE to find strategy with smaller mean square P\&L (smaller loss function).

\begin{figure}
\begin{subfigure}{.5\textwidth}
  \centering
  \includegraphics[width=.9\linewidth,height=.7\linewidth]{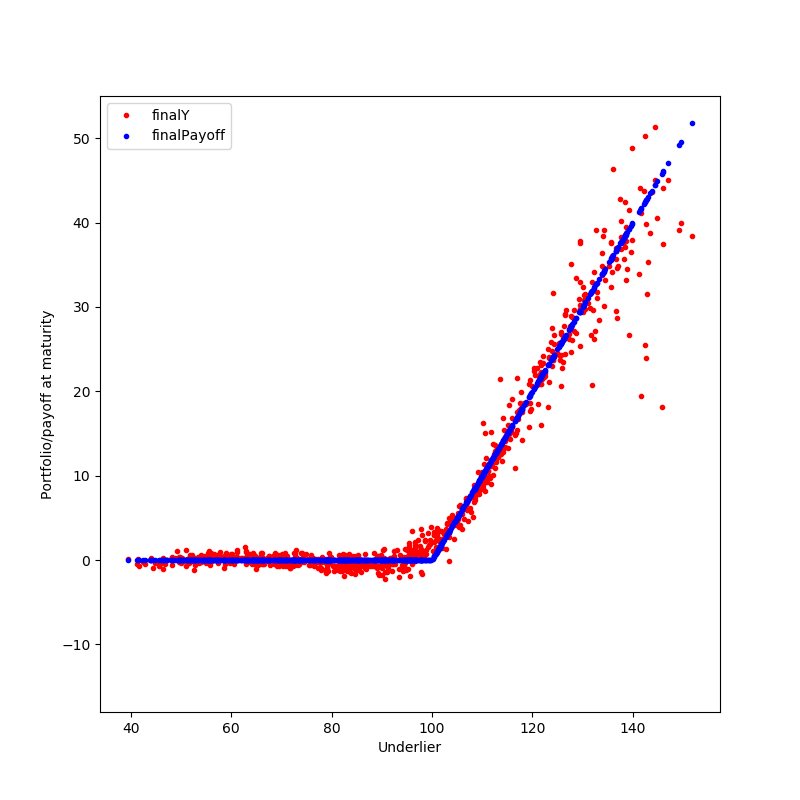}  
  \caption{At maturity.}
  \label{onedim50tsmaturityapprox}
\end{subfigure}
\begin{subfigure}{.5\textwidth}
  \centering
  \includegraphics[width=.9\linewidth,height=0.7\linewidth]{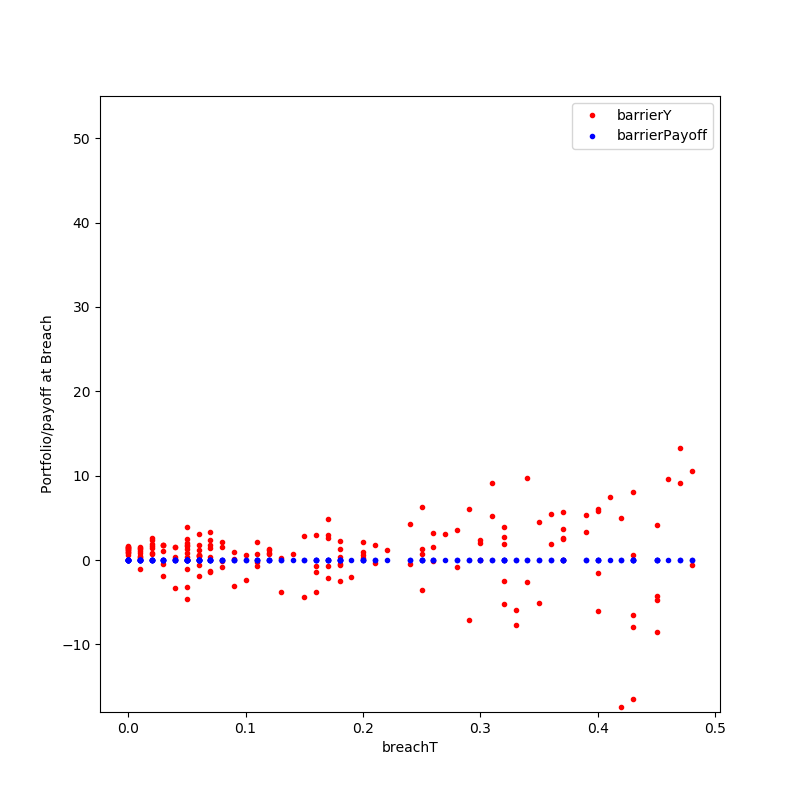}  
  \caption{At barrier.}
  \label{onedim50tsbarrierapprox}
\end{subfigure}
\caption[DeepBSDE for 1D Up-and-Out Call - replication at maturity
and barrier]{DeepBSDE:
replication of maturity and barrier payoff. One-dimensional case, 50 time
steps.}
\end{figure}

Figures \ref{onedim50tsmaturityapprox} and \ref{onedim50tsbarrierapprox} show
how well the DeepBSDE strategy replicates that payoff at maturity and the
knock-out barrier. 

Figures \ref{onedim50tsdeepbsdestrategy} and \ref{onedim50tsanalyticaldelta}
show histograms for the P\&L for the strategy computed by DeepBSDE and the
strategy given by the analytical delta of the analytical solution of the
continuous problem with the same barrier. One can see that the final P\&Ls span
a similar range and the analytical strategy peaks a bit higher, but otherwise
they have similar quantiles. The barrier P\&L for the analytical strategy is
asymmetric and mostly negative while the barrier P\&L for the DeepBSDE strategy
is more symmetric, leading to better P\&L behavior at
barrier breach.

\begin{figure}
\begin{subfigure}{.5\textwidth}
  \centering
  \includegraphics[width=.9\linewidth,height=.7\linewidth]{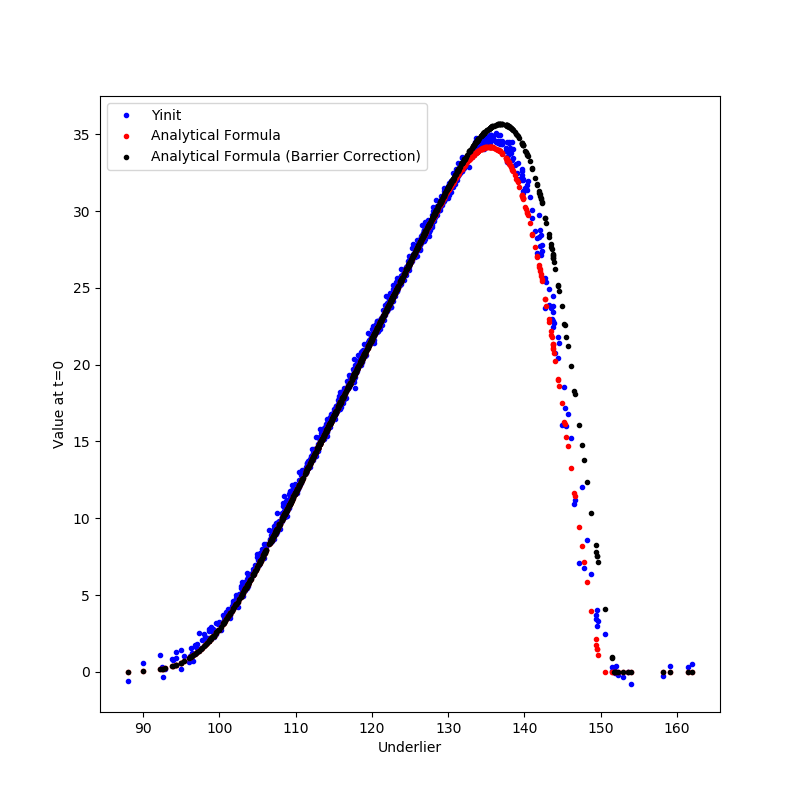}  
  \caption{50 time steps.}
  \label{tendim50ts}
\end{subfigure}
\begin{subfigure}{.5\textwidth}
  \centering
  \includegraphics[width=.9\linewidth,height=0.7\linewidth]{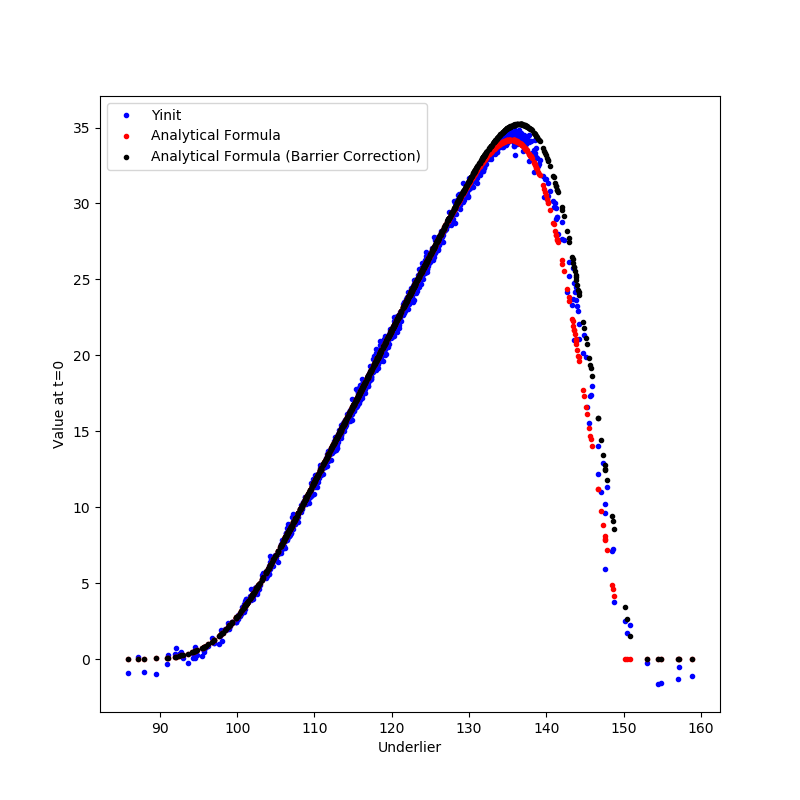}  
  \caption{100 time steps.}
  \label{tendim100ts}
\end{subfigure}

\begin{subfigure}{.5\textwidth}
  \centering
  \includegraphics[width=.9\linewidth,height=.7\linewidth]{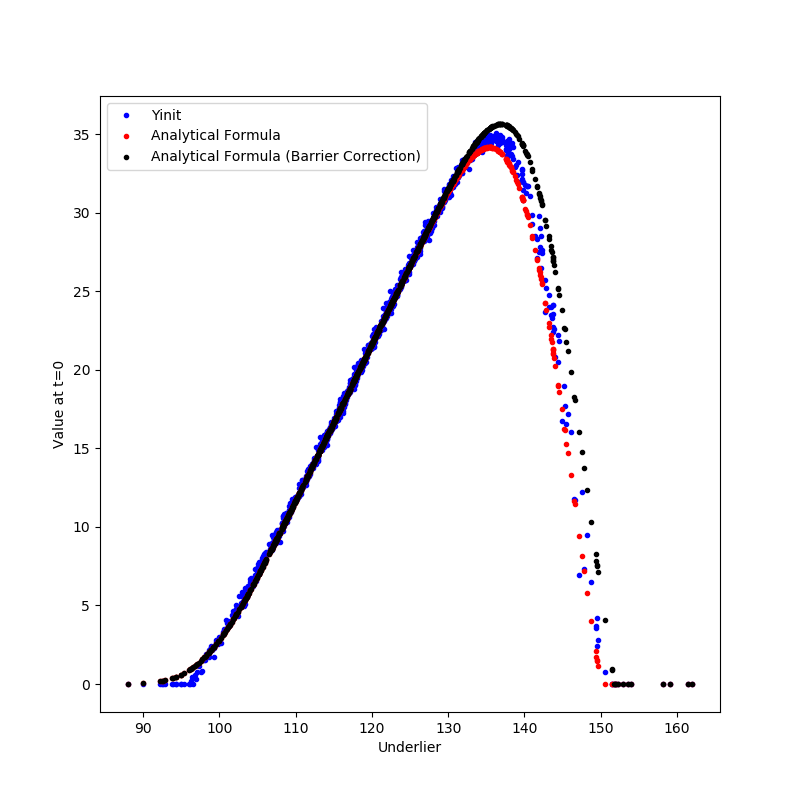}  
  \caption{50 time steps, positive.}
  \label{tendim50tspositive}
\end{subfigure}
\begin{subfigure}{.5\textwidth}
  \centering
  \includegraphics[width=.9\linewidth,height=0.7\linewidth]{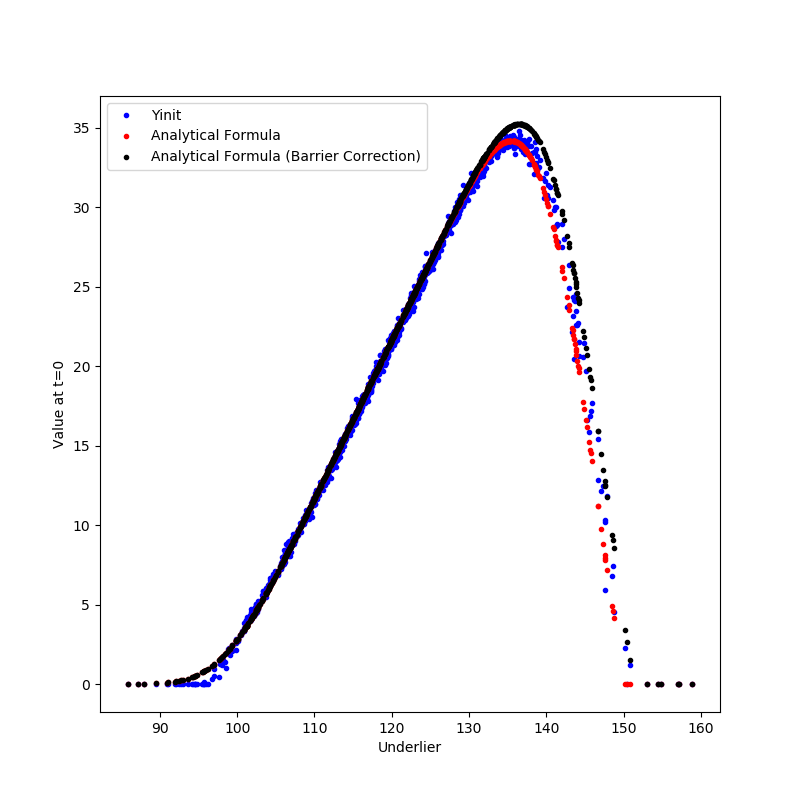}  
  \caption{100 time steps, positive.}
  \label{tendim100tspositive}
\end{subfigure}
\caption[DeepBSDE solution for 10D Up-and-Out Call - value]{DeepBSDE
value for ten-dimensional Up-and-Out Call on a geometric basket with different
 number of time steps (40 neurons in 2 hidden layers).}
\end{figure}

\begin{figure}
\begin{subfigure}{.5\textwidth}
  \centering
  \includegraphics[width=.9\linewidth,height=.7\linewidth]{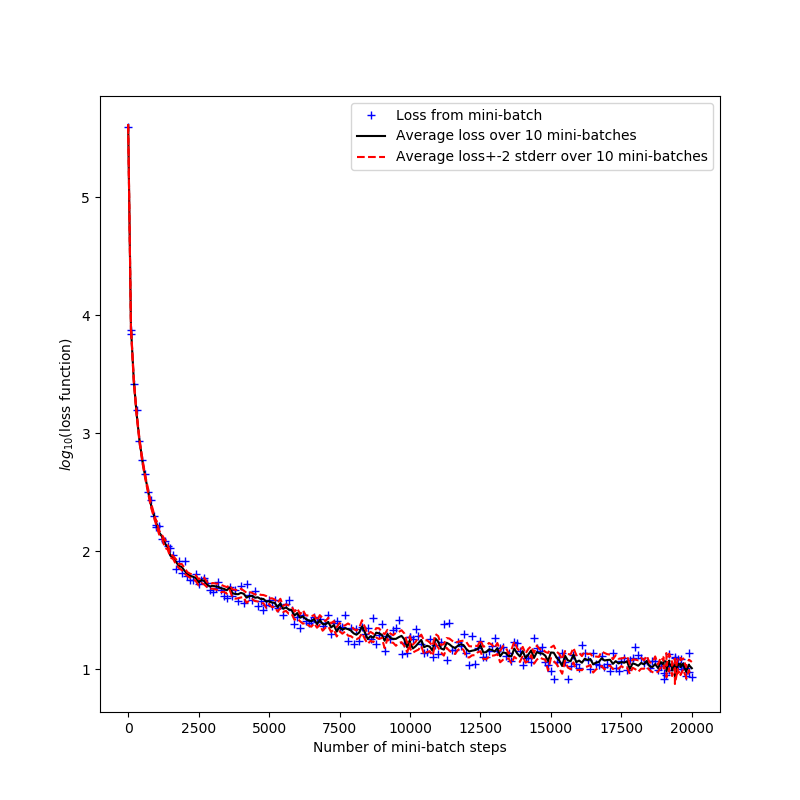}  
  \caption{Loss function.}
  \label{tendim100tsloss}
\end{subfigure}
\begin{subfigure}{.5\textwidth}
  \centering
  \includegraphics[width=.9\linewidth,height=0.7\linewidth]{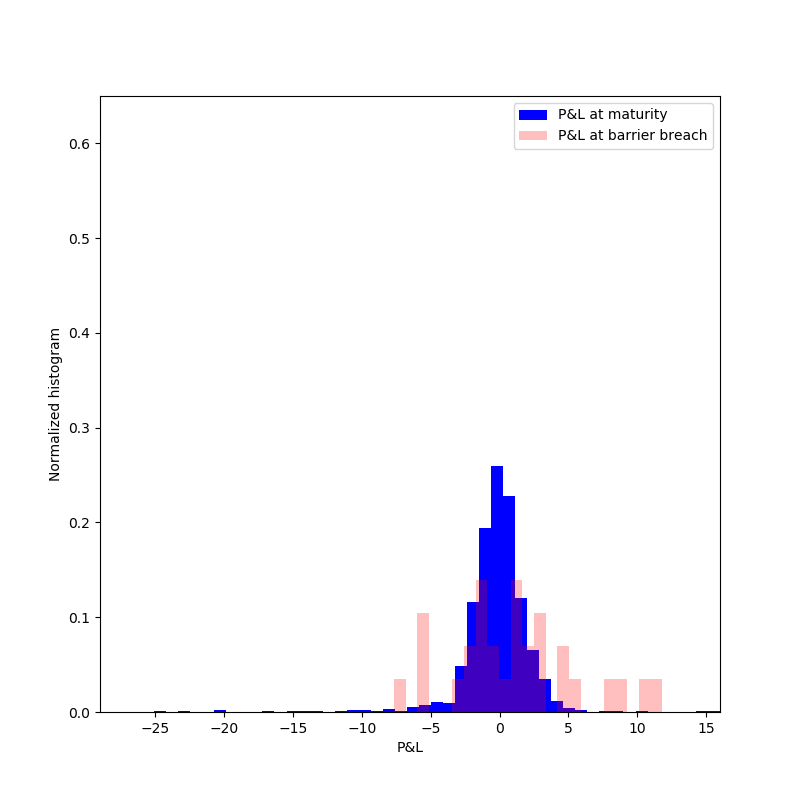}  
  \caption{Final and barrier P\&L.}
  \label{tendim100tspnl}
\end{subfigure}
\caption[DeepBSDE solution for 10D Up-and-Out Call - loss and P\&L]{DeepBSDE
loss function and final and barrier P\&L for ten-dimensional Up-and-Out
Call with 100 time steps (40 neurons in 2 hidden layers).}
\end{figure}

Figures \ref{tendim50ts}, \ref{tendim100ts}, \ref{tendim50tspositive}, and
\ref{tendim100tspositive} show the DeepBSDE solution for the ten-dimensional
example for different number of time steps. We are showing results both for a
standard fully connected DNN for $Y_0$ (top) and another architecture in which
positive part is applied to the output of the DNN. We projected points in $X$
space to the geometric basket underlier. Without being provided the geometric
mean as a feature, the trained DNN gives very similar results for different $X$
with the same geometric mean. The trained $Y_0$ function approximates the
analytical solution visually well. Close to the barrier, it is between the
analytical solution without and with continuity correction. We are not aware of
any other technique that would allow determining value function and strategy and
hedging performance for high-dimensional barrier options. Figure
\ref{tendim100tsloss} shows the evolution of the loss function during training
for 100 time steps case, showing convergence similar to the one-dimensional
case. Figure \ref{tendim100tspnl} shows that the final and barrier P\&L
histogram in ten dimensions looks similar to the one in one dimension,
demonstrating that the DeepBSDE method works similarly well in low and high
dimensions.

\section{Conclusion}
We proposed a novel deep neural network based computational graph architecture
to price Barrier Options via DeepBSDE, that captures whether the underliers'
price movement triggered the barrier or not in order to model the appropriate
payoff conditions and minimize the payoff error in order to learn the price of
the option at initial time and the hedging strategy at hedging times. We also
demonstrated the effectiveness of the hedging strategy via DeepBSDE learned
Deltas at discrete time instances.

\section{Acknowledgement}
The authors would like to thank Fernando Cela Diaz and Pallavi Abhang for their
help on setting up and running on   
distributed computing infrastructure, Orcan Ogetbil for discussions on barrier
option pricing, Vijayan Nair for discussion regarding methods, presentation, and
results and Agus Sudjianto for supporting this research.
The authors also would like to thank the two anonymous reviewers and the editors
for their comments that substantially improved the quality of the paper.
The authors report no conflicts of interest. The authors alone are responsible
for the content and writing of the paper.

\bibliographystyle{chicago_custom}
\bibliography{./reviewRefs}

\begin{thebibliography}{}

\bibitem[\protect\citeauthoryear{Chan-Wai-Nam, Mikael, and Warin}{Chan-Wai-Nam
  et~al.}{2019}]{chan2019machine}
Chan-Wai-Nam, Q., Mikael, J., and Warin, X. (2019).
\newblock Machine learning for semi linear {P}{D}{E}s.
\newblock {\em Journal of Scientific Computing\/}~{\em 79\/}(3), 1667--1712.
\newblock arXiv:1809.07609.

\bibitem[\protect\citeauthoryear{Cyr, Gulian, Patel, Perego, and Trask}{Cyr
  et~al.}{2019}]{cyr2019robust}
Cyr, E.~C., Gulian, M.~A., Patel, R.~G., Perego, M., and Trask, N.~A. (2019).
\newblock Robust training and initialization of deep neural networks: {A}n
  adaptive basis viewpoint.
\newblock {\em arXiv preprint arXiv:1912.04862\/}.

\bibitem[\protect\citeauthoryear{El~Karoui, Peng, and Quenez}{El~Karoui
  et~al.}{1997}]{el1997backward}
El~Karoui, N., Peng, S., and Quenez, M.~C. (1997).
\newblock Backward stochastic differential equations in finance.
\newblock {\em Mathematical finance\/}~{\em 7\/}(1), 1--71.
\newblock Also available on semanticscholar.org.

\bibitem[\protect\citeauthoryear{Han, Jentzen, and E}{Han
  et~al.}{2018}]{han2018solving}
Han, J., Jentzen, A., and E, W. (2018).
\newblock Solving high-dimensional partial differential equations using deep
  learning.
\newblock {\em Proceedings of the National Academy of Sciences\/}~{\em
  115\/}(34), 8505--8510.

\bibitem[\protect\citeauthoryear{Hientzsch}{Hientzsch}{2019}]{hientzsch2019intro}
Hientzsch, B. (2019).
\newblock Introduction to solving quant finance problems with time-stepped
  {F}{B}{S}{D}{E} and deep learning.
\newblock {\em arXiv preprint arXiv:1911.12231\/}.
\newblock Also available at SSRN: \url{https://ssrn.com/abstract=3494359} or
  \url{http://dx.doi.org/10.2139/ssrn.3494359}.

\bibitem[\protect\citeauthoryear{Kou}{Kou}{2003}]{kou2003pricing}
Kou, S. (2003).
\newblock On pricing of discrete barrier options.
\newblock {\em Statistica Sinica\/}~{\em 13}, 955--964.

\bibitem[\protect\citeauthoryear{Kremsner, Steinicke, and
  Sz\"{o}lgyenyi}{Kremsner et~al.}{2020}]{kremsner2020deep}
Kremsner, S., Steinicke, A., and Sz\"{o}lgyenyi, M. (2020).
\newblock A deep neural network algorithm for semilinear elliptic {P}{D}{E}s
  with applications in insurance mathematics.
\newblock {\em arXiv preprint arXiv:2010.15757v2\/}.

\bibitem[\protect\citeauthoryear{Lejay}{Lejay}{2002}]{lejay2002bsde}
Lejay, A. (2002).
\newblock B{S}{D}{E} driven by dirichlet process and semi-linear parabolic
  {P}{D}{E}. {A}pplication to homogenization.
\newblock {\em Stochastic processes and their applications\/}~{\em 97\/}(1),
  1--39.

\bibitem[\protect\citeauthoryear{Mercurio}{Mercurio}{2015}]{mercurio2015bergman}
Mercurio, F. (2015).
\newblock Bergman, {P}iterbarg, and beyond: {P}ricing derivatives under
  collateralization and differential rates.
\newblock In {\em Actuarial Sciences and Quantitative Finance}, pp.\  65--95.
  Springer.
\newblock Also available at SSRN: \url{https://ssrn.com/abstract=2326581} or
  \url{http://dx.doi.org/10.2139/ssrn.2326581}.

\bibitem[\protect\citeauthoryear{Pardoux}{Pardoux}{1998}]{pardoux1998backward}
Pardoux, {\'E}. (1998).
\newblock Backward stochastic differential equations and viscosity solutions of
  systems of semilinear parabolic and elliptic {P}{D}{E}s of second order.
\newblock In {\em Stochastic Analysis and Related Topics VI}, pp.\  79--127.
  Springer.

\bibitem[\protect\citeauthoryear{Perkowski}{Perkowski}{2010}]{perkowski2010}
Perkowski, N. (2010).
\newblock Backward {S}tochastic {D}ifferential {E}quations: {A}n introduction.
\newblock Available on semanticscholar.org.

\bibitem[\protect\citeauthoryear{Privault}{Privault}{2021}]{nprivaultonline2021}
Privault, N. (2021).
\newblock Notes on stochastic finance.
\newblock Lecture Notes. Available at
  \url{https://www.ntu.edu.sg/home/nprivault/indext.html}.

\bibitem[\protect\citeauthoryear{Raissi}{Raissi}{2018}]{raissi2018forward}
Raissi, M. (2018).
\newblock Forward-backward stochastic neural networks: {D}eep learning of
  high-dimensional partial differential equations.
\newblock {\em arXiv preprint arXiv:1804.07010\/}.

\bibitem[\protect\citeauthoryear{Shreve}{Shreve}{2004}]{book:shreve04}
Shreve, S. (2004).
\newblock {\em Stochastic Calculus for Finance II}.
\newblock Springer-Verlag New York.

\bibitem[\protect\citeauthoryear{Yu, Xing, and Sudjianto}{Yu
  et~al.}{2019}]{yu2019deep}
Yu, B., Xing, X., and Sudjianto, A. (2019).
\newblock Deep-learning based numerical {B}{S}{D}{E} method for {B}arrier
  options.
\newblock {\em arXiv preprint arXiv:1904.05921\/}.
\newblock Also available at SSRN: \url{https://ssrn.com/abstract=3366314} or
  \url{http://dx.doi.org/10.2139/ssrn.3366314}.

\end{thebibliography}
\end{document}